\documentclass[prx,twocolumn,floatfix,superscriptaddress,longbibliography,notitlepage]{revtex4-1}

\usepackage{graphicx, color, graphpap}
\usepackage{enumitem}
\usepackage{amssymb}
\usepackage{amsthm}
\usepackage{multirow}
\usepackage[colorlinks=true,citecolor=blue,linkcolor=magenta]{hyperref}
\usepackage[T1]{fontenc}
\usepackage{bbm}
\usepackage{thmtools,thm-restate}
\usepackage{verbatim}
\usepackage{mathtools}
\usepackage{titlesec}
\usepackage{amsmath}

\usepackage[caption = false]{subfig}

\newcommand{\bigzero}{\mbox{\normalfont\Large 0}}

\newcommand{\bigy}{\mbox{\normalfont Y}}

\long\def\ca#1\cb{} 

\newcommand{\ketbra}[2]{| \hspace{1pt} #1 \rangle \langle #2 \hspace{1pt} |}

\newcommand{\ket}[1]{|#1\rangle}               
\newcommand{\bra}[1]{\langle #1|}              
\newcommand{\dya}[1]{\ket{#1}\!\bra{#1}}
\newcommand{\ip}[2]{\langle #1|#2\rangle}      

\newcommand{\HC}{\mathcal{H}}

\newcommand{\RC}{\mathcal{R}}
\newcommand{\SC}{\mathcal{S}}

\newcommand{\XC}{\mathcal{X}}
\newcommand{\YC}{\mathcal{Y}}

\newcommand{\Tr}{{\rm Tr}}

\renewcommand{\geq}{\geqslant}
\renewcommand{\leq}{\leqslant}

\renewcommand{\Re}{\text{Re}}

\newcommand{\ad}{^\dagger}

\newcommand*{\id}{\openone}

{}
{}

\begin{document}
\title{Reformulation of the No-Free-Lunch Theorem for Entangled Data Sets}

\author{Kunal Sharma} 
\thanks{The first two authors contributed equally to this work.}
\address{Theoretical Division, Los Alamos National Laboratory, Los Alamos, New Mexico 87545, USA}
\address{Hearne Institute for Theoretical Physics and Department of Physics and Astronomy, Louisiana State University, Baton Rouge, Louisiana 70803, USA}

\author{M. Cerezo}
\thanks{The first two authors contributed equally to this work.}
\affiliation{Theoretical Division, Los Alamos National Laboratory, Los Alamos, New Mexico 87545, USA}
\affiliation{Center for Nonlinear Studies, Los Alamos National Laboratory, Los Alamos, New Mexico 87545, USA
}

\author{Zo\"{e} Holmes}
\address{Information Sciences, Los Alamos National Laboratory, Los Alamos, New Mexico 87545, USA}

\author{Lukasz Cincio}
\address{Theoretical Division, Los Alamos National Laboratory, Los Alamos, New Mexico 87545, USA}

\author{Andrew Sornborger}
\address{Information Sciences, Los Alamos National Laboratory, Los Alamos, New Mexico 87545, USA}

\author{Patrick J. Coles}
\affiliation{Theoretical Division, Los Alamos National Laboratory, Los Alamos, New Mexico 87545, USA}

\begin{abstract}
The no-free-lunch (NFL) theorem is a celebrated result in learning theory that limits one's ability to learn a function with a training data set. With the recent rise of quantum machine learning, it is natural to ask whether there is a quantum analog of the NFL theorem, which would restrict a quantum computer's ability to learn a unitary process with quantum training data. However, in the quantum setting, the training data can possess entanglement, a strong correlation with no classical analog. In this Letter, we show that entangled data sets lead to an apparent violation of the (classical) NFL theorem. This motivates a reformulation that accounts for the degree of entanglement in the training set. As our main result, we prove a quantum NFL theorem whereby the fundamental limit on the learnability of a unitary is reduced by entanglement. We employ Rigetti's quantum computer to test both the classical and quantum NFL theorems. Our Letter establishes that entanglement is a commodity in quantum machine learning.
\end{abstract}

\maketitle

{\it Introduction.---}There are very few fields of science and technology that have not been impacted by machine learning. Yet progress in machine learning has been anything but steady, with periods of stagnation interleaved with periods of advancement~\cite{haykin1994neural}. This reflects the deep and non-trivial nature of learning theory. In order to advance the theory, fundamental results needed to be proven on the trainability, expressibility, and scalability of learning architectures such as neural networks~\cite{rumelhart1986learning}.

One such fundamental result is the no-free-lunch (NFL) theorem~\cite{wolpert1997no,wolpert1995no,adam2019no,wolfnotess,shalev2014understanding}. At the conceptual level, the theorem states that different optimization procedures essentially perform the same when averaged over many problem instances and training data sets. At the mathematical level, the theorem has many alternative formulations, such as a statement that the average performance over all problem instances and training sets depends only on the size of the training data set and not on the optimization procedure. A consequence of this is that data must be considered the commodity or currency in machine learning that ultimately limits performance. Hence, this is why big data sets are viewed in such high regard.

Industry-built quantum computers of modest size are now publicly accessible over the cloud~\cite{steffen2016progress,karalekas2020quantum}. This raises the intriguing possibility of quantum-assisted machine learning, a paradigm that researchers suspect could be more powerful than traditional machine learning~\cite{biamonte2017quantum,dunjko2016quantum}. Various architectures for quantum neural networks (QNNs) have been proposed and implemented~\cite{schuld2014quest,schuld2014quantum,dunjko2018machine,verdon2018universal,farhi2018classification,ciliberto2018quantum,killoran2019continuous,cong2019quantum,beer2020training}. Some important results for quantum learning theory have already been obtained, particularly regarding the trainability~\cite{mcclean2018barren,cerezo2020cost,sharma2020trainability,grant2019initialization,verdon2019learning,volkoff2020large,skolik2020layerwise} and expressibility~\cite{sim2019expressibility} of QNNs for variational quantum algorithms \cite{VQE,bauer2016hybrid,mcclean2016theory,arrasmith2019variational,jones2019variational,Xiaosi,bravo-prieto2019,yuan2019theory,cirstoiu2019variational,cerezo2019variational,cerezo2020variational,QAQC,larose2019variational}. However, the scalability of QNNs (to scales that are classically inaccessible) remains an interesting open question.

A quantum version of the NFL theorem could play an important role in understanding the scalability of QNNs.
Recently, Poland et al.~\cite{poland2020no} made progress along these lines. They proved a lower bound on the average risk that depends only on the number of quantum states $t$ used for training. Here, the risk is the probability of incorrectly learning a unitary process, which is the natural quantum analog of the classical risk. Their bound tends to zero only as $t$ approaches the Hilbert-space dimension, which is exponentially large. This suggests that an exponentially large training data set is needed to learn a unitary. One can view this result as a roadblock in the path towards scaling QNNs, due to the apparent exponential (i.e., inefficient) scaling.

\begin{figure}[t]
    \centering
    \includegraphics[width=0.7\columnwidth]{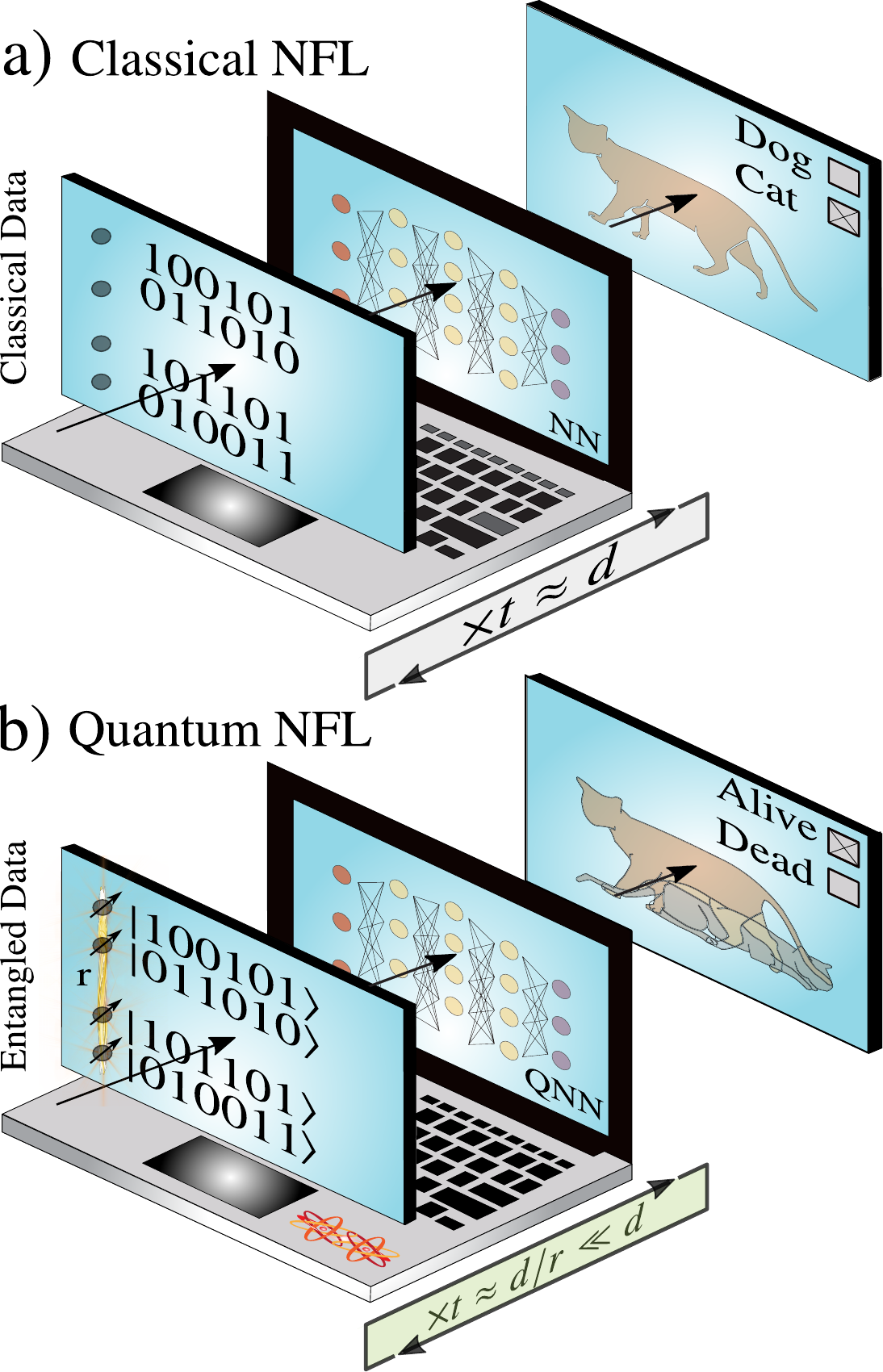}
    \caption{Depiction of the No-Free-Lunch setting. (a) In classical supervised learning, one employs training data of size $t$ to train a hypothesis to mimic the action of an unknown function on domain size $d$. Here we show input data in the form of bitstrings fed into a Neural Network (NN) to solve a binary classification problem. The NFL theorem indicates that it is the size of training data rather than the choice of optimization method that limits the average risk. Namely, small (large) $t$  leads to big (small) generalization errors on average.  (b) In quantum supervised learning, the goal is to learn a $d$-dimensional unitary process with $t$ quantum states serving as training data. For generality, we allow these states to possibly be entangled with a reference system, with the Schmidt rank $r$ quantifying the degree of entanglement. Here we show these states training a Quantum Neural Network (QNN) to classify quantum data (Schrodinger's cat being dead or alive). Our Quantum NFL theorem indicates that $r*t$ is the quantity that limits the average risk, and hence big entanglement (large $r$) leads to small generalization errors even when $t$ is small.}
    \label{fig:1}
\end{figure}

In this Letter, we consider a more general scenario, depicted in Fig.~\ref{fig:1}. Here, the goal is to learn a unitary with training data consisting of quantum states; however, these quantum states can now be entangled to a reference system. Such entangled states can be easily prepared on a quantum computer, and hence this scenario has practical relevance. A special case of this scenario is when the training data states have no entanglement with the reference system, corresponding to the scenario in Ref.~\cite{poland2020no}.

Our main result is a quantum NFL theorem that generalizes the result in Ref.~\cite{poland2020no} by allowing for an arbitrary amount of entanglement in the training data. An amazing feature of our theorem is that our lower bound on the average risk is reduced as the Schmidt rank $r$ of the entanglement grows. Furthermore the bound goes to zero when $r=d$, where $d$ is the Hilbert space dimension, regardless of the number of training data points $t$. Given that our bound is tight (i.e., it can be saturated), this implies that one does not need an exponentially large training data set in order to learn a unitary. Hence, our Letter establishes that both big data \textit{and} big entanglement are valuable in quantum machine learning, and that the currency of entanglement can lead to scalability.

Our Letter adds to the remarkable literature on entanglement as a resource. In communication theory, preshared entanglement allows one to transmit two bits of information by sending a single qubit~\cite{bennett1992communication}. In fundamental physics, an observer that is entangled to a system can guess the outcome of complementary measurements on that system, and this led researchers to generalize Heisenberg's uncertainty principle to allow for uncertainty reduction due to entanglement~\cite{berta2010uncertainty,berta2014entanglement,coles2017entropic}. Our Letter is analogous to these examples, albeit in a different context.

We note that in Ref.~\cite{Bisio10}, an important problem about learning an unknown unitary transformation from a finite number of examples was studied. In particular, Ref.~\cite{Bisio10} proved that whenever the unknown unitary is randomly drawn from a group the incoherent strategies achieve the ultimate performances for quantum learning. However, our results are different from Ref.~\cite{Bisio10} in the sense that we quantify the generalization error after training perfectly on the training set.

In what follows, we first discuss the classical NFL theorem. We then present our quantum NFL theorem, with the proof given in the Supplemental Material \footnote{See Supplemental Material at \href{http://link.aps.org/ supplemental/10.1103/PhysRevLett.128.070501}{http://link.aps.org/ supplemental/10.1103/PhysRevLett.128.070501} for more details, which contains Refs.~\cite{nielsen2002simple, collins2006integration, puchala2017symbolic, wolfnotess, poland2020no, singh2016average}}. Finally, we perform numerical tests of both NFL theorems. This includes an implementation on Rigetti's quantum computer, which allows us to effectively violate the classical NFL theorem and also verify our quantum NFL theorem. We note
that the Supplementary Material provides detailed proofs of all statements that follow.

\bigskip

{\it Results.---}
In classical supervised machine learning, NNFL arises in the setting depicted in Fig.~\ref{fig:1}(a).  Here the goal is to learn an unknown function $f$, where $f$ maps a discrete input set $\XC$ (of size $d_{\XC}$) to a discrete output set $\YC$ (of size $d_{\YC}$). In this setting one generates from $f$ a training  set $\SC$  in the form of $t$ ordered  input-output pairs as $\SC=\{(x_j,y_j):\, x_j\in \XC,y_j:=f(x_j)\in \YC \}_{j=1}^t$. This data is employed 
to train a hypothesis function $h_{\SC}$ such that it matches perfectly the action of  $f$ on the training data. The hope is that $h_{\SC}$ also makes accurate predictions on unknown, unseen data. However, as we will see, the NFL theorem provides a constraint on this.

To quantify how well the hypothesis function performs in predicting $f$ one defines the risk function $R_f(h_{\SC})$ as 
\begin{equation}\label{eq:risk}
    R_f(h_\SC)=\sum_{x\in \XC} \pi(x)\mathbb{P}\Big[f(x)\neq h_{\SC}(x)\Big].
\end{equation}
Specifically, $R_f(h_\SC)$ is the probability that $h_{\SC}(x)$ and $f(x)$ differ across $\XC$ when $x$ is sampled from the probability distribution $\pi(x)$.  While there are various mathematical versions of the NFL theorem~\cite{wolpert1997no,wolpert1995no,adam2019no,wolfnotess}, we follow the treatment in Ref.~\cite{wolfnotess}, which lower bounds the risk when averaged over training sets $\SC$ and functions $f$: 
\begin{equation}\label{eg:classicalNFL}
    \mathbb{E}_f[\mathbb{E}_\SC[R_f(h_\SC)]]\geq \left(1-\frac{1}{d_{\YC}}\right)\left(1-\frac{t}{d_{\XC}}\right)\,.
\end{equation}
This is an information-theoretic bound (and hence is independent of the optimization method employed in training), implying that the average risk is limited by the size of the training set $t$, with the bound going to zero if $t=d_{\XC}$. (Henceforth we drop the subscript when $d_{\XC} = d_{\YC} =d$, as in Fig.~\ref{fig:1}.)

As the NFL theorem is an information-theoretic result, the bound depends on the prior knowledge that one has about the set of maps from which $f$ is chosen. Given that we will ultimately be interested in unitary maps in the quantum setting, one can consider classical analogs of unitaries in the classical setting for a meaningful comparison. Hence, we reformulate the classical NFL theorem for both stochastic and bistochastic matrices, which are somewhat analogous to unitaries. In the Supplemental Material we show that the classical NFL theorem for stochastic and bistochastic matrices can be expressed as   \begin{align}\label{eq:StochNFL}
    \mathbb{E}_f[\mathbb{E}_S[R_f(h_{\SC})] \geq \left(1-\frac{t}{d}\right)F(d,t),
\end{align}
where $F(d,t)$ is the expectation over $f$ of the squared distance between $f(x)$ the $h_S(x)$.  In the stochastic case, we analytically find 
$F(d,t) = F(d) =  \frac{e^2(d-1)}{(d+1)d^{d+1}}\left((d-2)^{d+1}+2(d-1)^d \right) $. In the bistochastic case, we simplify the expression of $F(d,t)$ such that it can be numerically computed. The case of $f$ being a permutation matrix was considered in Ref.~\cite{poland2020no} and has a similar form as \eqref{eq:StochNFL}. All of these classical NFL results are conceptually similar, and dramatically different from the quantum case as we will see now. 

\bigskip

{\it Quantum NFL theorem.---}
Consider a quantum supervised learning task where the goal is to learn an unknown unitary $U$ that maps a $d$-dimensional input Hilbert space $\HC_\XC$ to a $d$-dimensional output Hilbert space $\HC_\YC$. Moreover, we consider a reference system $\RC$, with  $\HC_{\RC}$ denoting the associated Hilbert space, and we allow access to $\RC$ during the training process. We suppose that all training data states have the same Schmidt rank $r \in \{1,2,...,d\}$ across the cut $\HC_{\XC}\otimes \HC_{\RC}$. The training set is given by $t$ pairs of input-output states $\SC_Q = \{(\ket{\psi_j},\ket{\phi_j}):\,\ket{\psi_j}\in \HC_\XC\otimes \HC_{\RC},\ket{\phi_j}\in \HC_{\YC}\otimes \HC_{\RC} \}_{j=1}^{t}$. Here, the output states are given by $\ket{\phi_j}=(U\otimes \id_{\RC})\ket{\psi_j}$, where $\id_{\RC}$ is the identity over $\HC_{\RC}$.  During the training process, we allow for repeatable access to the states in $\SC_Q$. Perfect training corresponds to the condition where the hypothesis unitary $V_{\SC_Q}$ satisfies $|\ip{\tilde{\phi}_j}{\phi_j}|=1$ for all $j\in \{1,...,t\}$, where $\ket{\tilde{\phi}_j} =(V_{\SC_Q}\otimes \id_{\RC})\ket{\psi_j}$.

Similar to the classical case, we quantify the accuracy of the hypothesis $V_{\SC_Q}$ via the quantum risk function: 
\begin{equation}\label{eq:generrorQ}
R_U(V_{\SC_Q})=\int dx D_T^2(\dya{y},\dya{\tilde{y}}),
\end{equation}
defined as the average trace distance squared between the true output $\ket{y}=U\ket{x}$ and the hypothesis output  $\ket{\tilde{y}}=V_{S_{Q}}\ket{x}$, where $\ket{x} \in \HC_{\XC}$ and $\ket{y},\ket{\tilde{y}} \in \HC_{\YC}$. Here, $D_T(\rho,\sigma)=\frac{1}{2}||\rho-\sigma||_1$, and the integral is over the uniform Haar measure $dx$ on state space. Note that the risk is quantified on the smaller space $\HC_{\YC}$ while the training is performed on the larger space $\HC_{\YC}\otimes \HC_{\RC}$.

Averaging the risk $R_U(V_{\SC_Q})$  over all unitaries $U$ and training sets $\SC_Q$ leads to our main result:
\begin{align}\label{eq:g-qnfl}
    \mathbb{E}_U[\mathbb{E}_{\SC_Q}[R_U(V_{\SC_Q})] \geq 1 - \frac{r^2 t^2 + d + 1}{d(d+1)}\,, 
\end{align}
which is a NFL theorem for entanglement-assisted quantum supervised learning. The proof is presented in the Supplemental Material, where we also show that the bound in \eqref{eq:g-qnfl} can be stated more generally in that it holds for all choices of $\SC_Q$, and hence the average over $\SC_Q$ is trivial and can be removed from \eqref{eq:g-qnfl}. 
We show below in our numerical implementions that this bound is tight, and the inequality in \eqref{eq:g-qnfl} is saturated if the input states in $\mathcal{S}_Q$ are linearly independent (see Supplementary Material for more details).

Our proof for \eqref{eq:g-qnfl} relies on the assumption that the hypothesis unitary $V_{\mathcal{S}_Q}$ matches the target unitary $U$ perfectly on the training set. This condition reduces the unitary $U^{\dagger}V_{\mathcal{S}_Q}$ to a simple block diagonal form. We then employ the Weingarten calculus to calculate the average over all target uniaries, which reduces to \eqref{eq:g-qnfl}. We note that one does not need to perform tomography of states for evaluating the cost function. Rather, the overlap between the true output state and the output of the hypothesis unitary can be efficiently estimated, e.g., using the SWAP test.

\bigskip

{\it Implications of results.---}
Let us discuss the implications of \eqref{eq:g-qnfl}. First, consider the case of zero entanglement, $r=1$. In this case we recover the main result of Ref.~\cite{poland2020no}, which states that the average risk is non-zero when $t<d$ and can only go to zero when $t=d$. Typically, $d=2^n$ will be exponentially large in the quantum setting, with $n$ being the number of qubits, and hence this implies that an exponential amount of training data is needed to fully learn an unknown unitary.

At the other extreme, when there is maximal entanglement ($r=d$), one can see from~\eqref{eq:g-qnfl} that only one training pair is sufficient for the lower bound on the average risk to reach zero. In the language of quantum information theory~\cite{nielsen_chuang}, this single training data point corresponds to the ``Choi state'' of the target unitary $U$. More generally, \eqref{eq:g-qnfl} indicates that the key quantity is $r*t$. When $r*t$ is small (large), the bound on the average risk is high (low). Hence, even moderate amounts of entanglement can improve the performance of quantum machine learning, by reducing the training data requirements.

The standard goal of quantum algorithms is quantum speedup, which typically corresponds to complexity scaling polynomially in $n$, since classical algorithms often exhibit exponential scaling. Variational quantum algorithms, which train QNNs, are no exception, and any exponential scaling in such algorithms destroys quantum speedup. Consequently, the quantum NFL theorem of Ref.~\cite{poland2020no}, which corresponds to $r=1$ in our theorem, appeared to be a roadblock to quantum machine learning, since it suggested that an exponential amount of training data was required. Our Letter, on the other hand, appears to at least give some hope for quantum speedup with QNNs, provided that one has access to entangled training data. With that said, quantum speedup is a subtle issue, and we emphasize that \eqref{eq:g-qnfl} is derived under the assumption of perfect training. Hence one must analyze the complexity of training, and barren plateaus in training landscapes must be avoided in order to retain quantum speedup (see Discussion for elaboration).

In our implementations below, we compare the quantum and classical NFL theorems. We will argue that we observe an apparent violation of the classical NFL theorems. While these classical NFL theorems are of course valid under the setting of their formulation, this setting nevertheless does not allow for entangled data. Hence the apparent violation is due to the fact that the physical laws of nature allow for a more general setting than the assumed setting of these theorems. We also remark that one could allow for a reference system $\RC$ in the classical setting (like we do in the quantum setting). However, access to such a system would not change the bounds in the classical NFL theorems. This is because, in the classical setting, no correlation between $\RC$ and $\XC$ would be possible under the standard assumption that the joint state is a pure state. (Training with mixed states is not allowed since that would correspond to training with multiple pure states and, arguably, would be cheating.) Hence, allowing for $\RC$ in the classical setting is trivial.

\bigskip

{\it Implementations.---}
The availability of cloud-based quantum computers offers the possibility of testing the validity of NFL theorems with truly entangled data sets. In what follows, we present numerical results for quantum supervised learning, with the task of learning randomly generated unitaries, using entangled training states of increasing Schmidt rank. The details of our implementations are presented in the Supplemental Material.

\begin{figure}[t!]
    \centering
    \includegraphics[width=0.9\columnwidth]{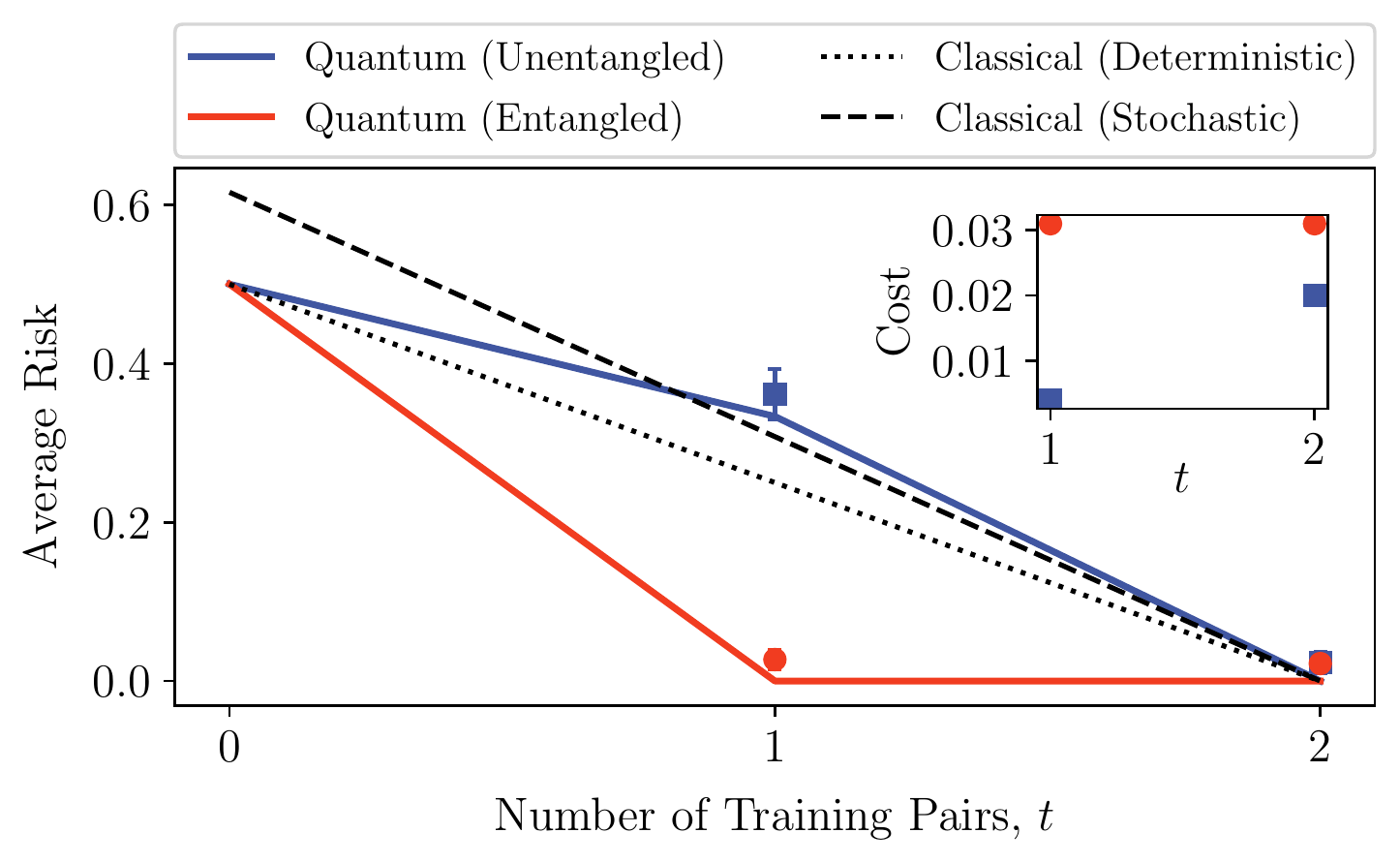}
    \caption{\textbf{Implementation on Quantum Hardware.} Here we plot the average risk after learning 10 single-qubit unitaries on the Rigetti Aspen-4 quantum computer using 10 training sets consisting of  $t=1,2$ unentangled $r=1$ (blue squares) and entangled $r=2$ (red circles) training states. The solid lines indicate the corresponding bounds imposed by our quantum NFL theorem, \eqref{eq:g-qnfl}. Note, that while the optimizations were performed on the quantum computer, the final risk $R_U(V_{\SC_Q})$ and optimal cost $C_U(V_{\SC_Q})$ (plotted in the inset and defined in the Supplemental Material) were calculated classically to allow an accurate (i.e., noiseless) evaluation of the success of the optimizations. In black we plot the classical deterministic (dotted) and stochastic (dashed) NFL theorems. }
    \label{fig:Rigetti}
\end{figure}


We first employ Rigetti's Aspen-4 quantum device~\cite{karalekas2020quantum} to learn $2\times 2$ unitaries. This involves a hybrid quantum-classical optimization loop where the quantum computer evaluates a cost function that quantifies the quality of the training on $\SC_Q$, and then the parameters of the hypothesis unitary are adjusted classically to reduce the cost. Figure~\ref{fig:Rigetti} shows the average risk versus $t$, after running this optimization loop, for training sets consisting of  $t=1,2$ unentangled ($r=1$) and entangled ($r=2$) states.  To compare the performance to the fundamental limits imposed by the NFL theorems, we also plot the classical bounds for deterministic \eqref{eg:classicalNFL} and stochastic \eqref{eq:StochNFL} maps as well as our quantum bound in \eqref{eq:g-qnfl}. Good agreement is observed for our quantum bound with the small discrepancies attributable to imperfect learning (due to the presence of quantum noise it was not possible to completely minimize the cost function as shown in the inset) and finite-size averaging when computing the average risk. The average risk using a single entangled training pair ($t=1, r=2$) is substantially lower than both the average risk using a single unentangled training pair ($t=1, r=1$) and that allowed by the deterministic and stochastic classical NFL theorems, suggesting an apparent violation of these classical bounds. 

\begin{figure}[t!]
    \centering
    \includegraphics[width=0.9\columnwidth]{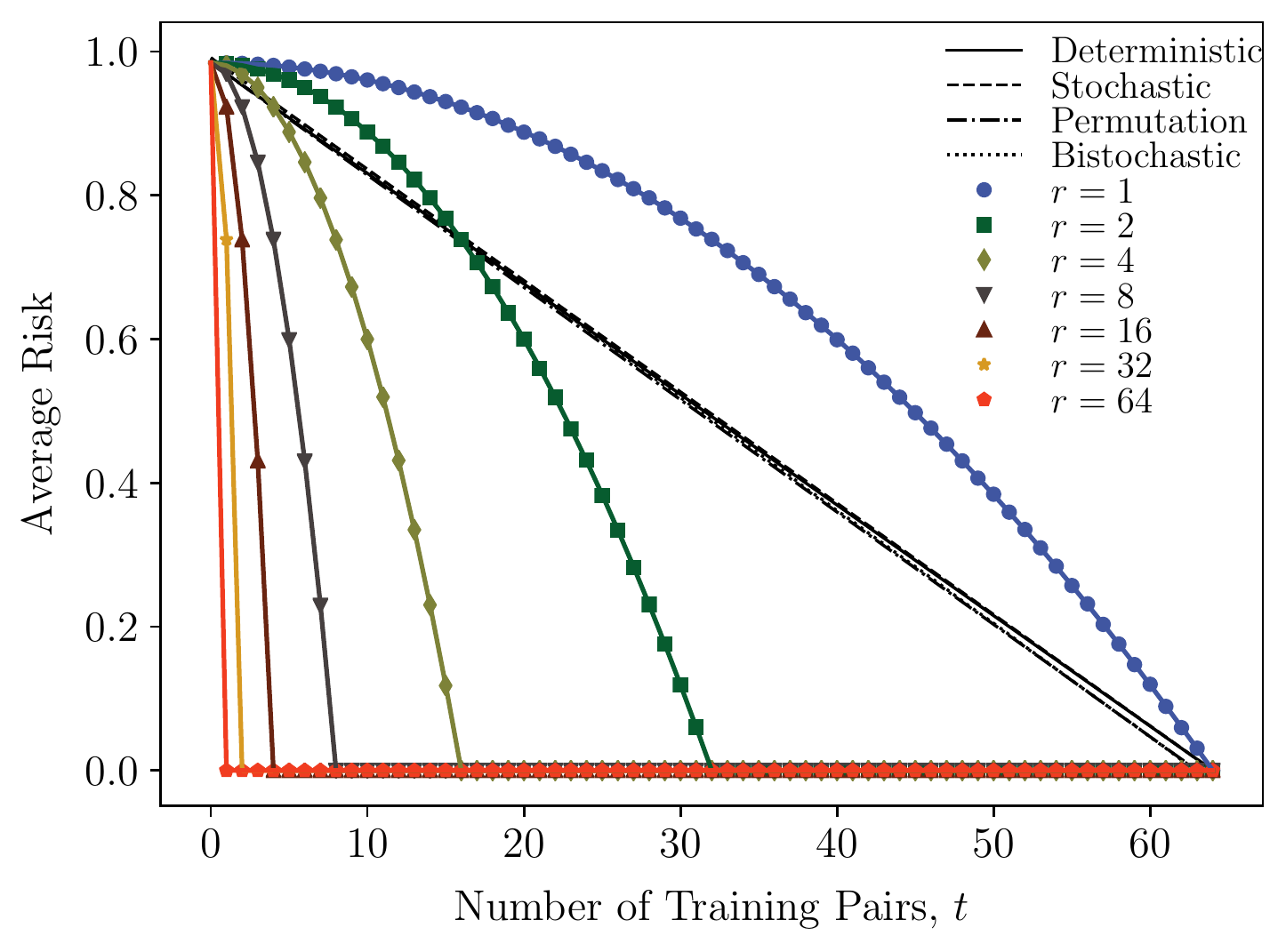}
    \caption{\textbf{Large-Scale Test of NFL Theorems.} We plot the average risk versus $t$ after learning 10 six-qubit unitaries on a simulator for 100 training sets. Each training set consisted of $t= 1,..., 64$ training pairs of rank $r =2^{0},..., 2^{6}$. The markers indicate the optimization results, whereas the solid lines indicate the bounds imposed by our quantum NFL theorem, \eqref{eq:g-qnfl}. The simulation error bars are $O(10^{-3})$ and therefore smaller than the size of the markers. In black, we plot the classical NFL bounds for deterministic (solid), stochastic (dashed), permutation (dot-dashed), and bistochastic maps.}
    \label{fig:Sim}
\end{figure}

While noise and other constraints limit the size of our quantum-hardware implementations, we can nevertheless explore larger systems on a simulator. Figure~\ref{fig:Sim} plots the average risk when learning 64-dimensional unitaries on a simulator for $t= 1,..., 64$ training states of Schmidt rank $r =2^{0},..., 2^{6}$. Near-perfect agreement between the simulation data and the bound in \eqref{eq:g-qnfl} is observed in all cases. Furthermore, for $r > 1$ it is possible to reduce the average risk below that allowed by four different classical NFL bounds (which have very similar behavior). We remark that 2-dimensional permutation and bistochastic matrices can be learned with a single training pair and hence it was not possible to violate the permutation and bistochastic classical bounds for the previous 2-dimensional implementation; whereas our 64-dimensional implementation easily violates these bounds.

\bigskip

{\it Discussion.---}
Quantum machine learning is a relatively new field that has already seen one major shift, from algorithms for the fault-tolerant era to variational methods for training Quantum Neural Networks (QNNs) in the near-term era. While several intriguing QNN architectures and training strategies have been proposed, rigorous results are urgently needed, in particular, to understand whether QNNs will offer a quantum speedup. In this Letter, we have contributed a rigorous theorem with implications for QNN scalability.  While it previously appeared that an exponentially large training set would be required to train a QNN, our quantum No-Free-Lunch (NFL) theorem shows that entanglement in the training data can compensate for and remove this exponential overhead. This suggests that entanglement should be considered as a valuable resource in reducing the generalization error in quantum machine learning. While our Letter provides a glimmer of hope that quantum machine learning could yield a quantum speedup (i.e., polynomial scaling), there are still several issues and open questions that we now discuss.

One potential issue is the complexity of obtaining the entangled training data in the first place. This complexity will depend on the mode of access to the data. 
We note that for the setting where a user has physical access to the target unitary, then it is advantageous to input a state entangled with a reference system to the unitary so that the user can generate input-output training data with entanglement~\cite{QAQC}. This procedure can overall decrease the average risk more efficiently in comparison to the input with no entanglement.

Another potential issue is the complexity of training. While our quantum NFL theorem assumes perfect training, it is possible that exponential scaling could be hidden in the training difficulty, especially in light of recent results on barren plateaus (exponentially vanishing gradients) in QNN cost function landscapes~\cite{mcclean2018barren,cerezo2020cost,sharma2020trainability}. While several promising strategies have been proposed to avoid barren plateaus in QNNs~\cite{grant2019initialization, verdon2019learning,volkoff2020large,skolik2020layerwise}, this remains an active area of research. We speculate that for cases when one needs only a polynomial number of shots for training (i.e., no barren plateau issues), learning a unitary using an entangled training set is more advantageous than training sets with no entanglement. Deriving a no-free-lunch (NFL) theorem that accounts for finite accuracy in training is an interesting open question that we leave for future work.

This highlights an important direction for future work. Naturally, it would be useful to extend the quantum NFL theorem to the case where one does not achieve perfect training on the training set. Such imperfect training could either be the result of shot noise or hardware noise, or could simply be due to local minima in the landscape. In this case, the lower bound in~\eqref{eq:g-qnfl} would not be saturated, and hence it would be of interest to tighten the bound to account for imperfect training.

\bigskip

K.S., L.C., and P.J.C. were supported by the U.S. Department of Energy (DOE), Office of Science, Office of Advanced Scientific Computing Research, under the Accelerated Research in Quantum Computing (ARQC) program. M.C. and P.J.C. were supported by the Laboratory Directed Research and Development program of Los Alamos National Laboratory (LANL) under project number 20180628ECR. M.C. was also supported by the Center for Nonlinear Studies at LANL. Z.P.H., A.T.S., and P.J.C. acknowledge support from the LANL ASC Beyond Moore's Law project.


\bibliography{ref.bib}






\onecolumngrid

\pagebreak

\appendix

\vspace{0.5in}

\setcounter{theorem}{0}

\begin{center}
	{\Large \bf Supplemental Material for {\it Reformulation of the No-Free-Lunch Theorem for Entangled Data Sets}}
\end{center}

We here present additional information and detailed proofs for the main results in the manuscript {\it Reformulation of the No-Free-Lunch Theorem for Entangled Data Sets}. First, in Section~\ref{sec:qnfl-sm} we provide a proof for the quantum No-Free-Lunch (NFL) theorem with entangled data sets. Then, in Section~\ref{sec:classicalNFL} we derive the classical NFL theorem bounds for the case when one is learning a deterministic process, and when one is learning probabilistic   processes (stochastic, and bistochastic). Finally, in Section~\ref{sec:ressource} we discuss what are the quantum resources needed to violate classical NFL theorems.

\section{Quantum No-Free-Lunch Theorem}\label{sec:qnfl-sm}

Let us first recall the notation required to derive the entanglement-assisted quantum NFL. Let $U$ denote the target unitary. Let $\HC_{\XC \RC} := \HC_{\XC} \otimes \HC_{\RC}$ and $\HC_{\YC \RC} := \HC_{\YC} \otimes \HC_{\RC}$ denote input and output Hilbert spaces, respectively, such that  $\dim(\HC_{\XC}) = \dim(\HC_{\YC}) = d$. Let $\SC_{Q}$ denote the training set of size $\vert \SC_Q \vert = t$, such that 
\begin{align}\label{eq:tdata-qnfl}
\SC_Q = \{(\ket{\psi_j},\ket{\phi_j}):\,\ket{\psi_j}\in \HC_{\XC\RC},\ket{\phi_j}\in \HC_{\YC \RC} \}_{j=1}^{t},
\end{align}
where 
\begin{equation}
    \ket{\phi_j} =(U \otimes \id_{\RC})\ket{\psi_j}~.
\end{equation}
 Moreover, we consider that all training data states have the same Schmidt rank $r \in \{1,2,...,d\}$ across the cut $\HC_{\XC}\otimes \HC_{\RC}$. 
Having perfectly trained a hypothesis unitary $V_{\SC_{Q}}$ on the training data states, we will have that 
 \begin{align}\label{eq:V-on-traindata}
\ket{\tilde{\phi}_j} := (V_{\SC_{Q}}\otimes \id_{\RC})\ket{\psi_j} =e^{i\theta_j} (U\otimes \id_{\RC})\ket{\psi_j}, ~ \forall \ket{\psi_j}\in \SC_{Q}.
\end{align}
Then to quantify the accuracy of the hypothesis $V_{\SC_{Q}}$, we define the risk function $R_U(V_{\SC_{Q}})$ as 
\begin{align}
    R_U(V_{\SC_Q}) &:= \int dx~ \frac{1}{4}\Vert U\ket{x}\bra{x}U\ad - V_{\SC_Q}\ket{x}\bra{x}V_{\SC_Q}\ad\Vert_1^2\\
    & = 1- \int dx~  \vert \langle x\vert U\ad V_{\SC_Q} \vert x \rangle \vert^2,
\end{align}
where $\ket{x}\in \HC_{\XC}$, and where the integral is over the uniform (Haar) measure $dx$ on state space, such that $\int dx =1$. Thus the risk function is proportional to average fidelity, which can also be expressed in terms of the Hilbert-Schmidt distance between $U$ and $V_{\SC_{Q}}$ as follows \cite{nielsen2002simple}:
\begin{equation}\label{eq:risk-sm}
R_U(V_{\SC_{Q}})=1-\frac{d+|\Tr[U\ad V_{\SC_{Q}}]|^2}{d(d+1)}\,.
\end{equation}

Let $W := U\ad V_{\SC_{Q}}.$ Then from the Schmidt decomposition of a pure state, each $\ket{\psi_j}$ can be represented as
 \begin{equation}
     \ket{\psi_j} = \sum_{k=1}^r \sqrt{c_{j,k}} \ket{\xi_{j,k}}_{\XC} \ket{\zeta_{j,k}}_{\RC}~,
 \end{equation} 
which implies that 
\begin{equation}
    \rho_{j} := \Tr_{\RC}[\ketbra{\psi_j}{\psi_j}] = \sum_{k=1}^r c_{j,k} \ketbra{\xi_{j,k}}{\xi_{j,k}}_{\XC}~,
\end{equation}
where $\sum_{k=1}^r c_{j,k} =1$. Then from \eqref{eq:V-on-traindata}, it follows that
\begin{align}
    e^{i \theta_j} 
    & = \Tr_{\XC \RC}[(W \otimes \id_{\RC})\ketbra{\psi_j}{\psi_j}] \label{eq:exp_thetaj}\\
    & = \Tr_{\XC}\left[ W \Tr_{\RC}[\ketbra{\psi_j}{\psi_j}]\right]\\
    & =  \sum_{k=1}^r c_{j,k} \Tr[W \ket{\xi_{j,k}}\bra{\xi_{j,k}}_{\XC}]\\
    & = \sum_{k=1}^r c_{j,k}  \beta_{j,k},  
\end{align}
where we defined
\begin{align}\label{eq:betajk}
     \beta_{j,k} =  \bra{\xi_{j,k}} W \ket{\xi_{j,k}}. 
\end{align}

By using the fact that $\vert e^{i \theta_j}\vert =1$ and $\sum_{k=1}^r c_{j,k} = 1$, we get that for each $j \in \{1, \dots, t\}$, 
\begin{align}\label{eq:betaj-expthj}
    \beta_{j,k} = e^{i \theta_j}, \forall k \in \{1, \dots, r\}.
\end{align}
We then have to consider three cases: (1) the states in $\SC_{Q}$ are orthogonal, (2) the states in $\SC_{Q}$ are non-orthogonal but linearly independent, and (3) the states in $\SC_{Q}$ are linear dependent. As we now show, all three cases lead to the same NFL bound.

If the input states in the set $\SC_{Q}$ are orthonormal then $\beta_{j,k}$ do not have to be the same for different values of $j$. Later we argue that if the set $\SC_{Q}$ is non-orthonormal (but still linear independent), then $\theta_j = \theta_l,~ \forall j,l \in \{1, \dots, t\}$. We now evaluate an average of the risk function $R_U(V_{\SC_Q})$ in \eqref{eq:risk-sm} over both $\SC_{Q}$ and $U$ when $\SC_{Q}$ is an orthonormal set.
Note that the $W$ is a unitary matrix that can be represented in the following form:
\[
W =\left(\begin{array}{@{}c|c@{}}
   \begin{matrix}
  e^{i\theta_1} & \dots & 0 \\
  \vdots & \ddots & \\
  0 &   & e^{i\theta_t}
  \end{matrix}
  & \bigzero \\
\hline
  \bigzero &
  \bigy\\
\end{array}\right),
\] 
which follows from \eqref{eq:betajk}, \eqref{eq:betaj-expthj}, and from the fact that $\sum_j \vert W_{ij}\vert^2 = \sum_i \vert W_{ij}\vert^2 =1 $. Here, $Y$ is a unitary matrix acting on a $(d-rt)$ dimensional Hilbert space, which is orthogonal to the space spanned by input states in $\SC_{Q}$, and there are $r$ copies of each of the $e^{i \theta_j}$ terms on the diagonal.

The order of the averages over the target unitaries $U$ and training sets $\SC_{Q}$ can be freely chosen given Fubini's theorem and so for convenience we first perform the averaging over $U$. Then the Haar average of term $|\Tr[U\ad V_{\SC_{Q}}]|^2$ in \eqref{eq:risk-sm} over $U$ is given by
\begin{align}
\int dU |\Tr[U\ad V_{\SC_{Q}}]|^2
& = \int dY \left[\left\vert r \bigg(\sum_{j=1}^{t} e^{i \theta_j}\bigg) + \Tr[Y]\right\vert^2\right] \label{eq:W-expanded}\\
  &= r^2\left(\bigg| \sum_{j=1}^t e^{i \theta_j}\bigg|^2\right) + \int dY \vert \Tr(Y) \vert^2 + \int dY 2r \Re \bigg[ \bigg(\sum_{j=1}^{t} e^{i \theta_j}\bigg)\Tr\big[Y\big]\bigg]\\
  & \leq r^2 t^2 + \int dY \vert\Tr(Y) \vert^2 + \int dY 2r \Re \bigg[ \bigg(\sum_{j=1}^{t} e^{i \theta_j}\bigg)\Tr\big[Y\big]\bigg] \label{eq:triangle-inequality}\\
  & = r^2 t^2 + 1. \label{eq:haar-integral}
 \end{align}
The first equality follows from the assumption that $Y$ is sufficiently random. We invoke the triangle inequality for the absolute value in \eqref{eq:triangle-inequality}. The first integral in \eqref{eq:triangle-inequality} can be calculated as follows: 
\begin{align}\label{eq:TrYsquaredInt}
    \int d\mu(Y) \vert \Tr[Y]\vert^2 &= \sum_{i,k} \int d\mu(Y) Y_{ii}Y^{*}_{kk} = \sum_{i,k} \frac{\delta_{ik}}{d-rt} = 1, 
\end{align}
where we used the fact that Haar integral over any unitary $V\in U(d)$, with $U(d)$ being the unitary group of degree $d$, satisfies the following property \cite{collins2006integration,puchala2017symbolic} :
\begin{align}\label{eq:sym_int1}
    \int dV v_{ij}v^*_{pk} = \frac{\delta_{ip}\delta_{jk}}{d},
\end{align}
where $v_{i j}$ are the matrix elements of $V$. 

Moreover, the second integral in \eqref{eq:triangle-inequality} is 
\begin{align}\label{eq:linearYzero}
    \int dY 2r \Re \bigg[ \bigg(\sum_{j=1}^{t} e^{i \theta_j}\bigg)\Tr\big[Y\big]\bigg] = 0,
\end{align}
which follows from the fact that the Haar measure is left- and right-invariant under the action of the unitary group of degree $d$ (in particular, under $-\mathbb{I}$). 

Since \eqref{eq:haar-integral} is independent of the set $\SC_{Q}$, the average of the risk function $R_U(V_{\SC_{Q}})$ over both $\SC_{Q}$ and $U$ reduces to 
\begin{align}\label{eq:Q-NFL}
\mathbb{E}_U[\mathbb{E}_{\SC_Q}[R_U(V_{\SC_{Q}})]] \geq   1 - \frac{r^2 t^2 + d + 1}{d(d+1)}.     
\end{align}

We now briefly argue that $\theta_j = \theta_l, \forall j, l\in\{1,\dots, t\}$ in \eqref{eq:betaj-expthj} if the input states in $\SC_{Q}$ are not orthonormal (but still linearly independent). 
Let $\ket{\psi_1}, \ket{\psi_2}\in \SC_{Q}$ be two non-orthogonal vectors. Then, without loss of generality $\ket{\psi_2}$ can always be represented as follows:
\begin{align}
    \ket{\psi_2} = c_1 \ket{\psi_1} + c_2 \ket{\psi_1^{\perp}},
\end{align}
where $\sum_i \vert c_i \vert^2 = 1$ and $|c_1|^2 \neq 0$. Then, from \eqref{eq:exp_thetaj}, we find that 
\begin{align}
    e^{i \theta_2} &= \langle \psi_2 \vert (W\otimes \id_{
    \RC}) \vert \psi_2\rangle\\
    & = \vert c_1 \vert^2 e^{i \theta_1} + \vert c_2 \vert^2 \langle \psi_1^{\perp} \vert (W \otimes \id_{\RC}) \vert \psi_1^{\perp}\rangle. 
\end{align}
Since $|e^{i \theta_2}|=1$, $\sum_i \vert c_i \vert^2 = 1$ and $|c_1|^2 \neq 0$, the aforementioned equation is satisfied if and only if 
\begin{align}\langle \psi_1^{\perp}\vert (I_R \otimes  W_A) \vert \psi_1^{\perp} \rangle = e^{i \theta_2} = e^{\theta_1},
\end{align}
which implies that $\theta_1 = \theta_2$. This procedure can then be recursively applied  to the rest of the states in the training set, which leads to  $\theta_j = \theta_k, \forall j, k\in\{1,\dots, t\}$. 
Therefore, $\vert \Tr[U\ad V_{\SC_Q}]\vert^2$ in \eqref{eq:W-expanded} reduces to
\begin{align}\label{eq:TrWsquared}
 \vert \Tr[U\ad V_{\SC_Q}]\vert^2 =   \vert \Tr[W]\vert^2 = r^2 t^2 + \vert \Tr[Y]\vert^2 + 2 r t\Re\left[\Tr[Y]\right],
\end{align}
and hence, assuming the input states in $\SC_{Q}$ are linearly independent but non-orthogonal, the bound in \eqref{eq:Q-NFL} is saturated.

Finally, we note that if $\SC_{Q}$ contains input states that are linearly dependent, then $t$ in \eqref{eq:Q-NFL} gets replaced by $\widetilde{t}$, where $\widetilde{t}$ denotes the effective number of linearly-independent states in the set $\SC_{Q}$. 

Bringing all three of these cases together, we obtain our quantum NFL theorem in \eqref{eq:Q-NFL}. Moreover, since the bound in \eqref{eq:Q-NFL} holds for all choices of $\SC_Q$, the average over $\SC_Q$ is trivial and can be removed from $\eqref{eq:Q-NFL}$.

\subsection{Fluctuations in the quantum risk}

In this section we derive an expression for the variance in the risk function $R_U(V_{\SC_Q})$ as in \eqref{eq:risk-sm} over all unitaries and training sets, i.e. 
\begin{equation}\label{eq:FluctDef}
   \sigma_{R}^2 =  \mathbb{E}_U\left[\mathbb{E}_{\SC_Q}\left[[R_U(V_{\SC_Q})]^2\right]\right] - \left( \mathbb{E}_U\left[\mathbb{E}_{\SC_Q}[R_U(V_{\SC_Q})]\right]\right)^2  \ .
\end{equation}
For simplicity, and to align with our numerical implementations, we consider the case when the training set is composed of states which are  linearly independent but non-orthonormal. In this case, as discussed in Section \ref{sec:qnfl-sm}, it is possible to perfectly learn the target unitary $U$ on the subspace spanned by the training set, up to a global phase $e^{i \theta}$. Therefore, $W = U\ad V_{\SC_Q}$ reduces to
\[
W = \left(\begin{array}{@{}c|c@{}}
   \begin{matrix}
  e^{i \theta} & \dots & 0 \\
  \vdots & \ddots & \\
  0 &   & e^{i \theta}
  \end{matrix}
  & \bigzero \\
\hline
  \bigzero &
  \bigy\\
\end{array}\right) \, .
\] 
Note that for $r*t  = d$ we know that $W = e^{i \theta} \id$, which implies that $\sigma_{R}^2 =0$.
However, for $r*t  < d$, the risk $R_U(V_{\SC_Q})$, following \eqref{eq:TrWsquared}, evaluates to 
\begin{equation}\label{eq:UnAveragedRiskQNFL}
   R_U(V_{\SC_Q}) = \frac{d}{d+1} - \frac{1}{d(d+1)} \left(  r^2 t^2 + \vert\Tr(Y) \vert^2 + 2r t \Re \left[ \Tr\left[Y\right] e^{i \theta} \right] \right).
\end{equation}
Then, from Section \ref{sec:qnfl-sm}, the average of $R_U(V_{\SC_Q})$ over all unitaries and training sets is given by 
\begin{align}\label{eq:AverageRiskQNFL}
    \mathbb{E}_U[\mathbb{E}_{\SC_Q}[R_U(V_{\SC_Q})]] = 1 - \frac{r^2 t^2 + d + 1}{d(d+1)} \, .
\end{align}
Substituting \eqref{eq:UnAveragedRiskQNFL} and \eqref{eq:AverageRiskQNFL} into \eqref{eq:FluctDef}, we find that
\begin{equation}\label{eq:variance-expanded}
   \left(\sigma_{R}\right)^2 = \frac{1}{d^2 (d+1)^2} \int  d\mu(Y) \left(  \vert\Tr(Y) \vert^4   + 4 r^2 t^2  \Re( \Tr(Y) e^{i \theta} )^2  + 4r t  \vert\Tr(Y) \vert^2 \Re(\Tr(Y) e^{i \theta})  -1  \right)
\end{equation}
where we have simplified the expression using \eqref{eq:TrYsquaredInt} and \eqref{eq:linearYzero}.

Let $V\in U(d)$. Then by invoking the following formula for symbolic integration with respect to the Haar measure on a unitary group \cite{collins2006integration,puchala2017symbolic} 
\begin{align}\label{eq:sym_int2}
    \int d\mu(V)v_{i_1j_1}v_{i_2j_2}v_{i_1'j_1'}^{*}v_{i_2'j_2'}^{*}&=\frac{\delta_{i_1i_1'}\delta_{i_2i_2'}\delta_{j_1j_1'}\delta_{j_2j_2'}+\delta_{i_1i_2'}\delta_{i_2i_1'}\delta_{j_1j_2'}\delta_{j_2j_1'}}{d^2-1}
-\frac{\delta_{i_1i_1'}\delta_{i_2i_2'}\delta_{j_1j_2'}\delta_{j_2j_1'}+\delta_{i_1i_2'}\delta_{i_2i_1'}\delta_{j_1j_1'}\delta_{j_2j_2'}}{d(d^2-1)},
\end{align}
the first integral in \eqref{eq:variance-expanded} reduces to
\begin{align}
    \int  d\mu(Y) \vert\Tr(Y) \vert^4 &= \sum_{i,k} \int d\mu(Y) Y_{ii} Y_{jj} Y^{*}_{kk} Y^{*}_{ll} \\
    & =  \frac{2}{(d-rt)^2-1} \sum_{ik} \delta_{ik} \sum_{jl} \delta_{jl} - \frac{2}{((d-rt)^2-1)(d-rt)} \sum_{ij} \delta_{ij} = 2.
\end{align}

The second integral in \eqref{eq:variance-expanded} can be evaluated as follows:
\begin{equation}
  \int  d\mu(Y) \Re(\Tr(Y) e^{i \phi} )^2 =  \frac{1}{2} \int d\mu(Y) \Re(\Tr(Y)^2 e^{ 2 i \phi} ) + \frac{1}{2} \int d\mu(Y) \vert \Tr(Y) \vert^2  = \frac{1}{2}
\end{equation}
where we used \eqref{eq:TrYsquaredInt} to show that 
\begin{align}
    \int d\mu(Y) \vert \Tr(Y) \vert^2 = 1,
\end{align}
and from the left- and right-invariance of the Haar measure under the unitary $-i \id$, it follows that 
\begin{align}
\frac{1}{2} \int d\mu(Y) \Re(\Tr(Y)^2 e^{2 i \phi} )=0.
\end{align} 

\begin{figure}[t!]
    \centering
    \includegraphics[width=0.4\columnwidth]{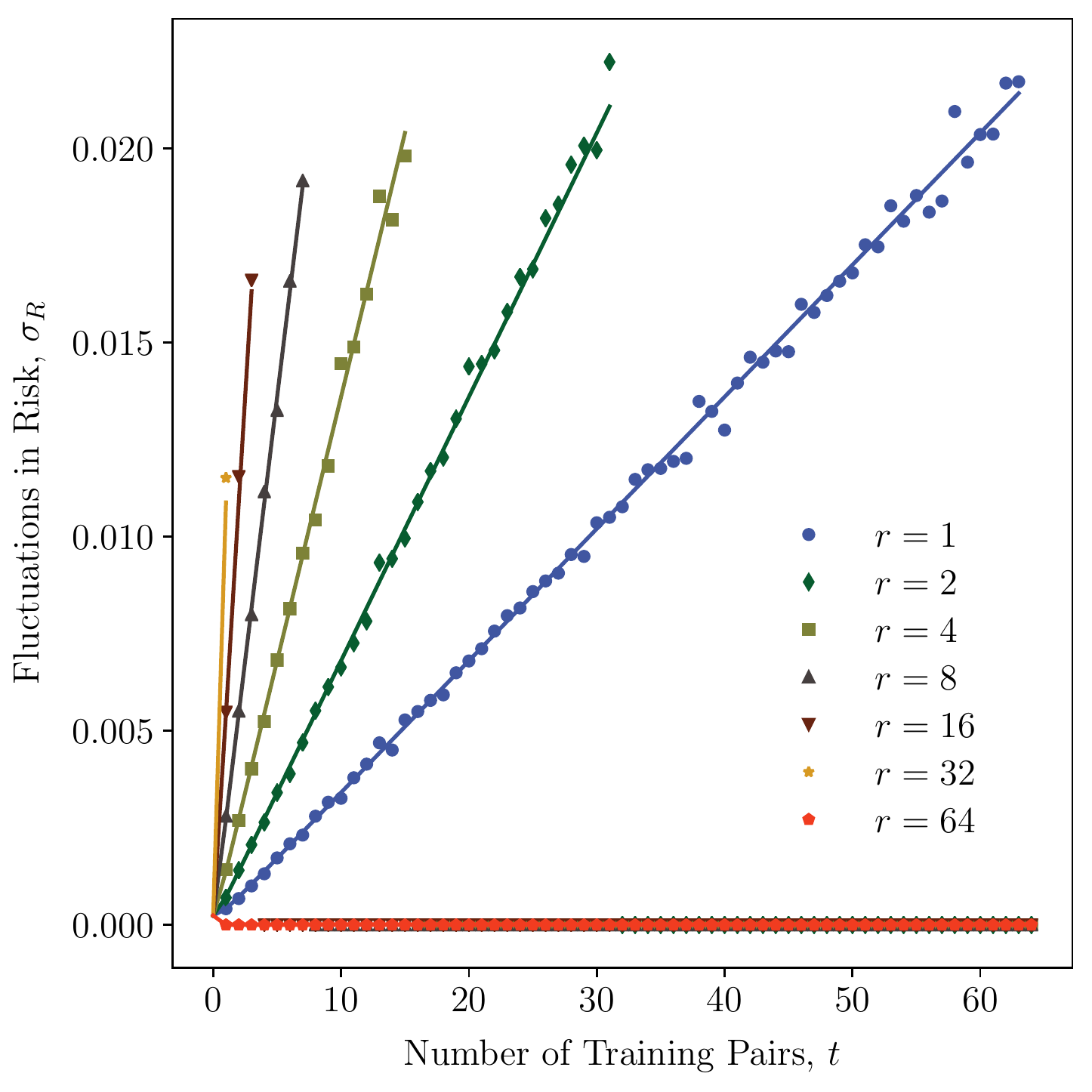}
    \caption{\textbf{Fluctuations in Risk.} The figure plots  the standard deviation in the risk after learning a six qubit unitary on a simulator for $t= 1$ to $t= 64$ training states of rank $r =2^0$ to $r = 2^6$. The markers indicate the optimisation results whereas the solid lines the predicted fluctuations according to \eqref{eq:Fluct}}
    \label{fig:SimFluct}
\end{figure}

Similarly, the third integral in \eqref{eq:variance-expanded} vanishes due to the left- and right-invariance of the Haar measure under the unitary $- \id$. Thus the standard deviation in the risk is given by 
\begin{equation}\label{eq:Fluct}
\begin{aligned}
& \sigma_{R} =
\left\{
\begin{array}{ll}
\frac{\sqrt{2 r^2 t^2 + 1}}{d(d+1)}  & \ \ \ \text{if } r*t < d \\
0 &  \ \ \ \text{otherwise} \, .
\end{array} 
\right.
\end{aligned}
\end{equation}
As shown in Fig.~\ref{fig:SimFluct}, this expression matches well with the data obtained from the 6 qubit implementation on a numerical simulator. Moreover, in the limit of large dimensions the standard deviation in the risk scales as $\approx (r* t)/d^2$ for $1 \ll r*t < d$, which implies that for high values of $d$ the fluctuations in the risk are exponentially suppressed. 

\section{Details on Implementations}

\begin{figure}[t!]
    \centering
        \includegraphics[width=0.5\columnwidth]{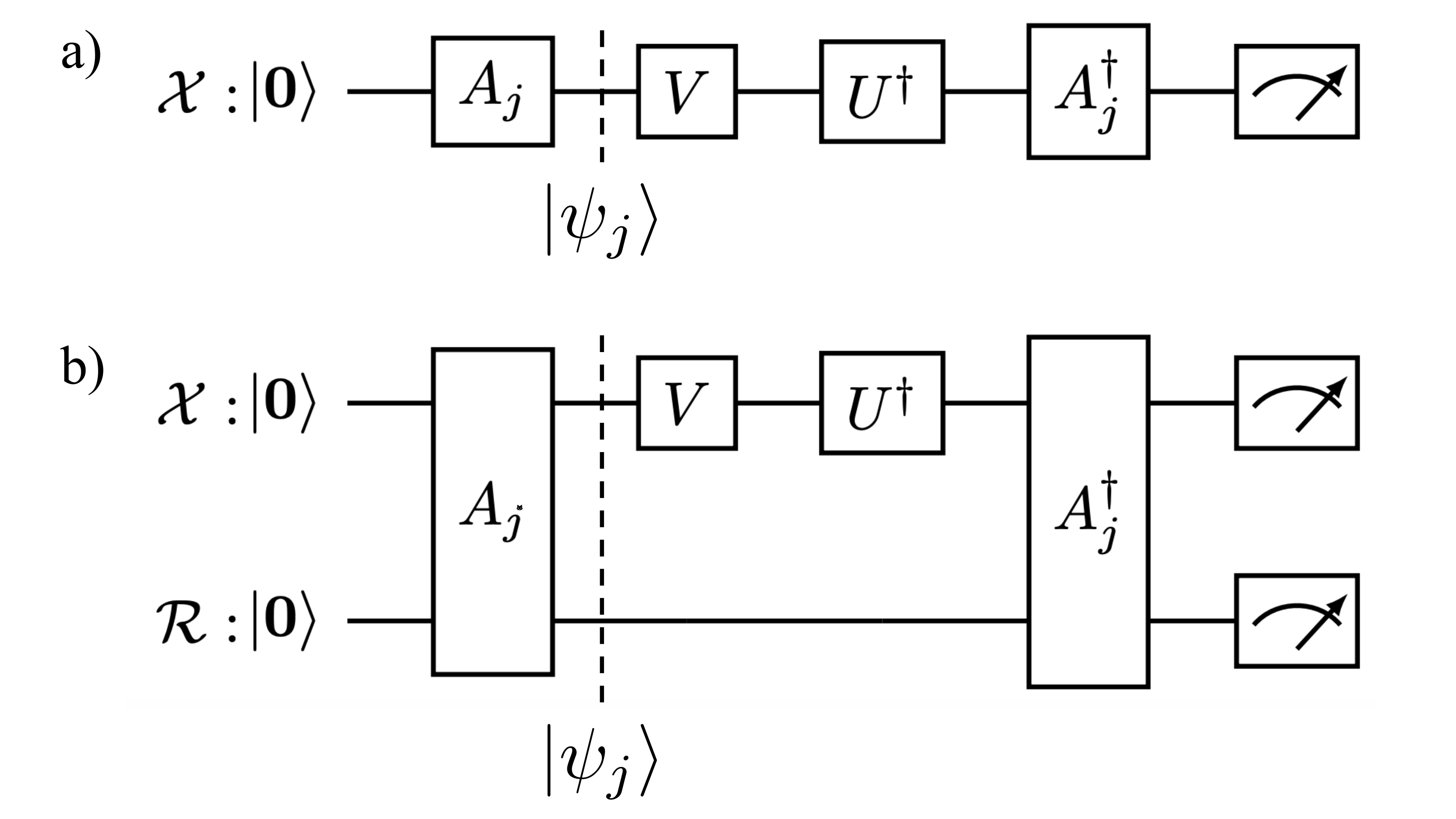}
    \caption{\textbf{Circuit for Testing NFL Theorem.} Here we show the circuits used to evaluate the cost function in~\eqref{eq:CostFunc}. Panels (a) and (b) correspond to the cases when the system $\XC$ is unentangled or entangled, respectively, with the reference $\RC$. Here $A_j$ denotes a circuit that efficiently prepares the state $\ket{\psi_j}$ from the all-zero state $\ket{\mathbf{0}}$, i.e. $\ket{\psi_j} = A_j \ket{\mathbf{0}}$. The upper and lower wires respectively denote the system and reference qubits in $\XC$ and in $\RC$. We remark that the probability of measuring the all-zero state is equal to the state overlap $|\bra{\phi_j} ( V \otimes \mathbb{I} ) \ket{\psi_j}|^2$. For the implementation on Rigetti's quantum computer the cost was evaluated from 1000 shots.}
    \label{fig:circuit}
\end{figure}

Here we elaborate on the methods used in our numerical implementations.
For both implementations we first generated a Haar random unitary $U$ and a random training set $\SC_Q$ consisting of $t$ pairs of training states of rank $r$. To learn $U$ we found the optimal hypothesis unitary $V_{\SC_Q}$ by minimizing the cost function 
\begin{equation}\label{eq:CostFunc}
    C_U(V) = 1 -  \frac{1}{t} \sum_{j=1}^{t} | \bra{\phi_j} ( V \otimes \mathbb{I} ) \ket{\psi_j}|^2 \, .
\end{equation}
This cost function quantifies the overlap between an input training state evolved under the hypothesis unitary, $ ( V \otimes \mathbb{I} ) \ket{\psi_j}$, and the output training state, $\ket{\phi_j}$, averaged over the $t$ training pairs. 
The circuits used to measure the state overlap are shown in Fig.~\ref{fig:circuit}, with the unentangled and entangled cases shown in panels (a) and (b) respectively. 

Having obtained the optimal hypothesis $V_{\SC_Q}$, we calculated the risk $R_U(V_{\SC_Q})$ defined in \eqref{eq:risk-sm}.
The average was calculated over 10 random unitaries and 10 random training sets for the 2-dimensional implementation on the Rigetti quantum computer, and over 10 random unitaries and 100 random training sets in the case of the 64-dimensional implementation on the simulator.

\section{Classical No-Free-Lunch Theorems}\label{sec:classicalNFL}

\subsection{Classical No-Free-Lunch Theorem For Deterministic Functions}

In this section we provide a simple proof for the classical NFL theorem, where we follow the treatment \cite{wolfnotess}. Later we provide an alternate proof. 
We begin by recalling the notation used throughout the proof. 
Let $\XC$ and $\YC$ denote discrete input and output sets of sizes $d_{\XC}$ and $d_{\YC}$, respectively. Let $f:\XC\rightarrow \YC$ be an unknown function from $\XC$ to $\YC$, and let $\SC$ denote a training set in the form of ordered input-output pairs
\begin{align}\SC=\{(x_j,y_j):\,x_j\in \XC,y_j := f(x_j)\in \YC \}_{j=1}^t. 
\end{align}
The goal  is to train a hypothesis function $h_{\SC}$ from the training set $\SC$ to guess the function $f$. That is, we employ a learning algorithm  to model the data
so that 
\begin{equation}
    h_{\SC}(x_j)=y_j=f(x_j), \forall (x_j,y_j)\in \SC~.
\end{equation}
Moreover, to quantify the performance of a hypothesis $h_{\SC}$, we define the    risk function as     
\begin{align}
R_f(h_\SC)=\sum_{x\in \XC} \pi(x)\mathbb{P}\Big[f(x)\neq h_{\SC}(x)\Big],
\end{align}
where $x$ is sampled from $\XC$ with respect to a distribution $\pi(x)$. Assuming that $\pi(x)$ is the uniform distribution, and $d_{\XC}=d_{\YC}=d$, we get
\begin{align}\label{eq:classDetNFL}
    \mathbb{E}_f[\mathbb{E}_{\SC}[R_f(h_{\SC})]] &= \frac{1}{d}\mathbb{E}_{f}\left[\mathbb{E}_{\SC}\left[\sum_{x \in \SC}\mathbb{P}\Big[f(x)\neq h_{\SC}(x)\Big]\right]\right]+\frac{1}{d}\mathbb{E}_{f}\left[\mathbb{E}_{\SC}\left[\sum_{x \notin \SC}\mathbb{P}\Big[f(x)\neq h_{\SC}(x)\Big]\right]\right]\\
    & \geq  \frac{1}{d}\mathbb{E}_{f}\left[\mathbb{E}_{\SC}\left[\sum_{x \notin \SC}\mathbb{P}\Big[f(x)\neq h_{\SC}(x)\Big]\right]\right]\\
    & = \frac{1}{d}\left[\left(d- t \right)\left(1- \frac{1}{d}\right) \right] \\
    &= \left(1 - \frac{t}{d} \right) \left(1 - \frac{1}{d} \right) \label{eg:classicalNFLsupp}.
\end{align}
The first inequality follows from the assumption that $h_{\SC}(x)= f(x)$ when $x \in \SC$. However, if $x\not\in \SC$, then $h(x)$ is completely random. In other words, $h(x)$ can take, with equal probability, any value in $\YC$. Therefore, if one were to simply guess at random, then the error probability is $(1-1/d)$. Moreover, the  coefficient $(d-t)$ in the second equality arises from the fact that there are $(d - t)$ data points that are not in  $\SC$.  

Note that the above derivation of the classical NFL theorem defines the risk as the \textit{probability} that the  hypothesis function outputs an incorrect bitstring. However to derive a quantum NFL we define the risk in terms of the \textit{trace distance squared} between the outputs of the hypothesis and the and target functions. The latter will be helpful to put the classical and quantum bounds on an equal footing. As we now show, the classically NFL theorem can also be derived by defining the risk in terms of a squared $1$-norm distance. 

Specifically, consider the risk function defined as the average trace distance squared between the output of the hypothesis $h_{\SC}$ and target $f$ functions on a random input $x$, sampled with respect to $\pi(x)$. That is, let
\begin{align}
    R_f(h_{\SC}) = \sum_{x\in \XC} \pi(x) \left(\frac{1}{2}\Vert f(x) - h_{\SC}(x) \Vert_1\right)^2 \, .
\end{align}
Here, $\frac{1}{2}\Vert a - b \Vert_1$ denotes the $1$-norm distance between two vectors $a$ and $b$. Then, let us assume again that $\pi(x)$ is the uniform distribution. Without loss of generality, we can also assume that both $\XC$ and $\YC$ are sets (of equal size $d$) consisting of $d$-dimensional bitstrings with Hamming weight one. Then, consider the following chain of inequalities: 
\begin{align}\label{eq:classDetNFLv2}
   \mathbb{E}_f[ R_f(h_{\SC}) ]
    & \geq \frac{1}{d} \sum_{x \notin \SC} \mathbb{E}_f\left[\left(\frac{1}{2}\Vert f(x) - h_{\SC}(x) \Vert_1\right)^2\right] \\
    & = \frac{ (d-t)}{d} \mathbb{E}_f\left[\left(\frac{1}{2}\Vert f(x) - h_{\SC}(x) \Vert_1\right)^2\right] \label{eq:deterministic-eq1}  \\
    & \geq \frac{(d-t)}{d} \mathbb{E}_f\left[\left(\frac{1}{2}\Vert f(x) - h^{\text{opt}}_{\SC}(x) \Vert_1\right)^2\right] \label{eq:deterministic-eq2} \\
    & = \frac{(d-t)}{d} \frac{1}{d} \left(d - 1\right)\\ \label{eq:deterministic-eq3}
    &  = \left(1-\frac{t}{d}\right) \left(1- \frac{1}{d} \right).
\end{align}
The first inequality follows from the fact that if we assume perfect training, then $h_{\SC}(x)$ guesses correctly $f(x)$ for all $x\in \SC$. The second equality follows from the fact that  the average distance $\Vert f(x) - h_{\SC}(x) \Vert_1^2$  should be same for all $x\not\in \SC$. The second inequality holds by definition, as $h_{\SC}^{\text{opt}}$ is the optimal hypothesis which minimizes the trace-distance square in \eqref{eq:deterministic-eq1}.
In this case, the deterministic optimal hypothesis corresponds to randomly guessing (with equal probability) a $d$-dimensional bit string of Hamming weight one. Since there are $d$ possible guesses, then we will have that $(d-1)$ times the distance will be $\frac{1}{2}\Vert f(x) - h_{\SC}^{\text{opt}}(x)\Vert_1 =1$. 

While \eqref{eq:classDetNFL}, or equivalently \eqref{eq:classDetNFLv2}, is valid for general functions $f$, we recall that the No-Free-Lunch Theorem is an information theoretic result, meaning that the right-hand-side  of~\eqref{eg:classicalNFLsupp} can change when specializing $f$ to a specific set of maps. It was shown in \cite{poland2020no}, that if $f$ is known to be an invertible function the risk is bounded as
\begin{equation}\label{eq:inverse-func}
    \mathbb{E}_f[\mathbb{E}_\SC[R_f(h_\SC)]]\geq 1-\frac{t+1}{d} \, .
\end{equation}
Let us here remark that, as expected, for finite $t$ and in the $d\rightarrow \infty$ limit, the average risks in Eqs.~\eqref{eq:deterministic-eq3} and~\eqref{eq:inverse-func} goes to one.

\subsection{Classical NFL theorems for probabilistic maps}

The standard classical no-free-lunch theorem in \eqref{eg:classicalNFLsupp} corresponds to the task of learning a deterministic matrix, where every element is either one or zero and each column sums to one. Such matrices, as the name suggests, represent deterministic processes where a bit string of Hamming weight one is mapped to another bit string of Hamming weight one. Similarly, the NFL for invertible deterministic processes, \eqref{eq:inverse-func}, corresponds to the task of learning a permutation matrix, a deterministic matrix in which not only the columns but also the rows sum to one. 

However, since in the quantum case one wishes to learn a unitary matrices (which can quantify probabilistic processes) in this section we derive NFL theorems for more general classes of matrices, which will allow for a fairer comparison between the quantum and classical NFL theorems. Namely, we consider here stochastic and doubly-stochastic matrices, which can be used to model classical probabilistic processes.  Stochastic matrices are matrices in which each element is real and positive and every column sums to one. Doubly-stochastic matrices are the subset of stochastic matrices, 
such that every row (as well as every column) sums to one.

\begin{figure}[t!]
    \centering
    \subfloat[Relationship between classical bounds]{
    \includegraphics[width=0.4\columnwidth]{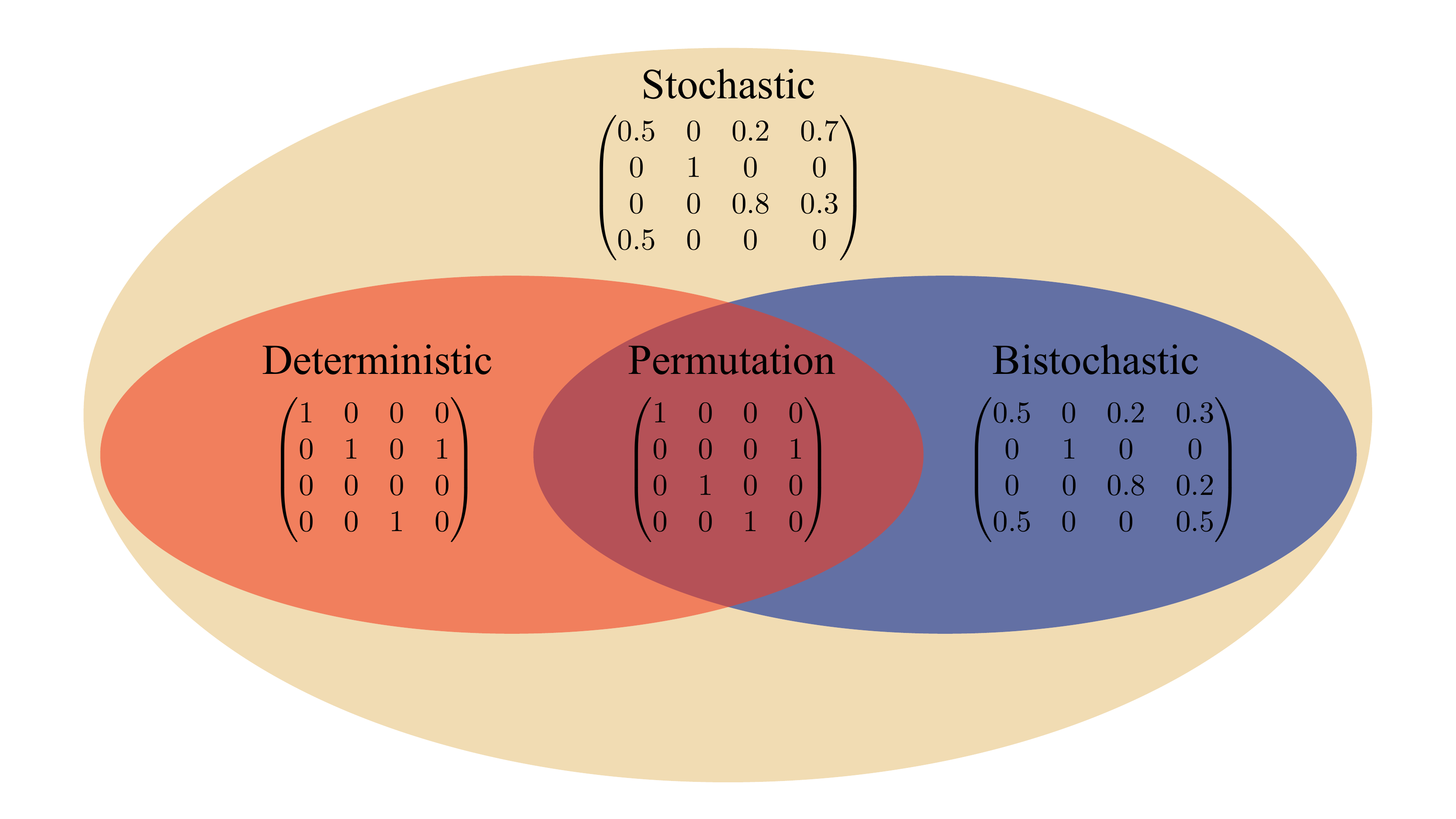}} \hspace{1cm}
    \subfloat[Rank required to violate classical bound]{\includegraphics[width=0.4\columnwidth]{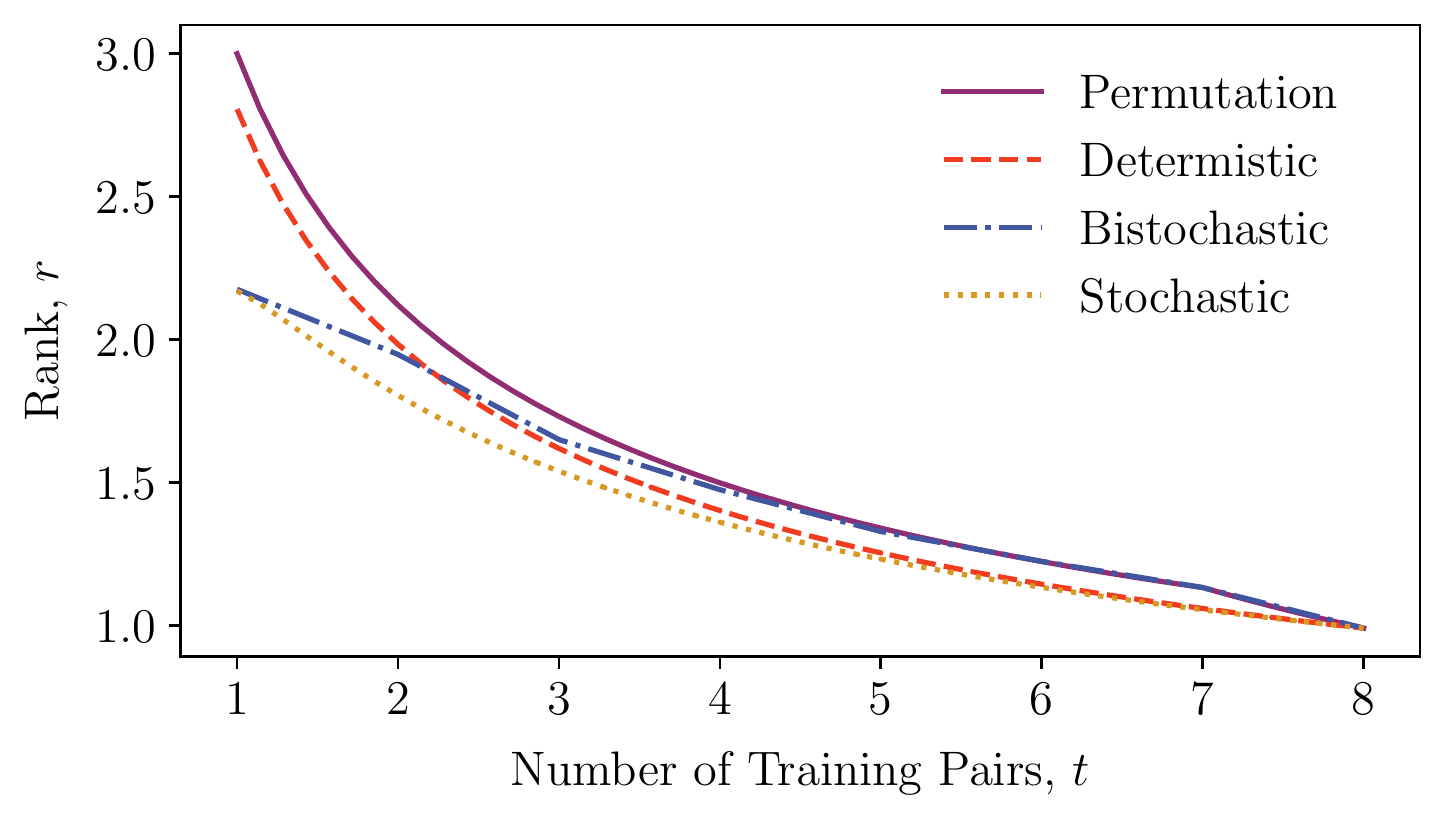}}
    \caption{\textbf{Classical bounds.} (a) A Venn diagram indicating the overlapping sets of stochastic, deterministic, bistochastic and Permutation matrices along with an example of each sort. Stochastic matrices are matrices for which each element is real and positive and every column sums to one. Deterministic matrices are stochastic matrices where every element is either 1 or 0 and bistochastic matrices are stochastic matrices where every row sums to one. Permutation matrices are matrices which are both deterministic and bistochastic. Hence, stochastic, deterministic, bistochastic and permutation matrices form a partial order with permutation matrices more constrained that bistochastic and deterministic matrices which in turn are more constrained than stochastic matrices. We therefore expect more resources to be required to learn stochastic matrices than permutation matrices, with the resources required to learn bistochastic and deterministic matrices sitting between the two. This is indeed confirmed in (b) where we plot the rank $r$ of training states required to violate the stochastic, deterministic, bistochastic and permutation NFL theorems as a function of the number of training pairs $t$ when learning a $8\times 8$ dimensional matrix. The number of pairs $t$ required to violate the permutation bound is greater than the rank required to violate the stochastic bound, with the rank required to violate the deterministic and bistochastic bounds sitting between the two.}
    \label{fig:ClassicalBounds}
\end{figure}

As indicated in Fig.~\ref{fig:ClassicalBounds}(a), stochastic, deterministic, bistochastic, and permutation matrices form a partial order where permutation matrices are more constrained than bistochastic and deterministic matrices, which in turn are more constrained than stochastic matrices. From an information theoretic perspective, the more prior knowledge we have about the matrix (or maps) to be learnt, the less resources we should need to learn it. We therefore expect more resources to be required to learn stochastic matrices than permutation matrices, with the resources required to learn bistochastic and deterministic matrices sitting between the two extremes. 

\subsubsection{Classical NFL theorem for stochastic matrices}\label{sec:stochastic-nfl}

In this section, we derive a No-Free-Lunch theorem for learning stochastic matrices.  A stochastic matrix is a square matrix, such that each entry is non-negative and the sum of the entries of each column is equal to one. We formulate the task of learning a stochastic matrix in terms of learning its columns. Hence, let us assume that $M_{\text{S}}$ is a $d\times d$ stochastic matrix. Let $\XC$ and $\YC$ denote sets of $d$-dimensional bitstrings of Hamming weight one, and let $|\XC| = |\YC| = d$. Let $f:\XC \to \mathcal{C}(\YC)$ be a map corresponding to the unknown stochastic matrix $M_{\text{S}}$, taking bitstrings from $\XC$ to a convex combination of bitstrings in $\YC$. Here we can define the training set  $\SC$  as
\begin{align}\label{eq:tset_stochastic}
 \SC=\{(x_j,y_j):\,x_j\in \XC,y_j = f(x_j)\in \mathcal{C}(\YC) \}_{j=1}^t.
\end{align}
Moreover, similar to previous sections, the goal is to train a hypothesis map $h_{\SC}$ from the training set $\SC$ to estimate the target map $f$, such that 
\begin{align}\label{eq:perfect_train_stochastic}
    h_{\SC}(x_j) = y_j = f(x_j),~ \forall (x_j, y_j) \in \SC. 
\end{align}
To assess the performance of the learning task, we define the risk function as follows:
\begin{align}\label{eq:1norm}
    R_f(h_{\SC}) := \frac{1}{d}\sum_{x \in \XC} \left(\frac{1}{2}\Vert f(x) - h_{\SC}(x) \Vert_1\right)^2,
\end{align}
where $\frac{1}{2}\Vert a - b \Vert_1$ denotes the $1$-norm distance between two vectors $a$ and $b$. Moreover, we recall that we define the risk such that in the limit $d\rightarrow \infty$, the average risk should be one. 

We now find the average of the risk over all maps $f$. Consider the following chain of inequalities: 
\begin{align}
   \mathbb{E}_f [R_f(h_{\SC})] &= \frac{1}{d} \sum_{x \in \SC}  \mathbb{E}_f\left[\left(\frac{1}{2}\Vert f(x) - h_{\SC}(x) \Vert_1\right)^2\right] + \frac{1}{d} \sum_{x \notin \SC}  \mathbb{E}_f\left[\left(\frac{1}{2}\Vert f(x) - h_{\SC}(x) \Vert_1\right)^2\right] \label{eq:error_train_test_stochastic}\\
    & \geq  \frac{1}{d} \sum_{x \notin \SC}  \mathbb{E}_f\left[\left(\frac{1}{2}\Vert f(x) - h_{\SC}(x) \Vert_1\right)^2\right]\\
    & =   \frac{(d-t)}{d} \mathbb{E}_f\left[\left(\frac{1}{2}\Vert f(x) - h_{\SC}(x) \Vert_1\right)^2\right] \label{eq:avg-risk-stoch-eq1} \\
       & \geq \frac{(d-t)}{d}\left[ \min_{h_{\SC}(x)} \bigg(\mathbb{E}_f\left[ \left(\frac{1}{2}\Vert f(x) - h_{\SC}(x) \Vert_1\right)^2\right] \bigg)\right] \label{eq:minimization}\\
           & =\frac{(d-t)}{d} \left[\mathbb{E}_f\left(\frac{1}{2}\Vert f(x) - h^{\text{opt}}_{\SC}(x) \Vert_1\right)^2\right] \label{eq:average_expression}\\
    & = (1-t/d) F(d),
\end{align}
where 
\begin{align}
 F(d) := \mathbb{E}_f\left(\frac{1}{2}\Vert f(x) - h^{\text{opt}}_{\SC}(x) \Vert_1\right)^2 =    \frac{e^2(d-1)}{(d+1)d^{d+1}}\left((d-2)^{d+1}+2(d-1)^d \right) . \label{eq:stochastic_finalexp}
\end{align}

The first inequality follows from the assumption that the hypothesis map $h_{\SC}$ predicts the target map $f$ on the training data perfectly. Since the action of the map $f$ is unknown on all $x \notin \SC$, the average of $\Vert f(x) - h_{\SC}(x) \Vert^2_1$ over all functions should be the same for all $x \notin \SC$. Thus we get the second equality, where $x \notin \SC$. The second inequality follows as we minimize over all hypothesis maps. As we discuss below,  the expression in \eqref{eq:minimization} is minimized by
\begin{align}\label{eq:hsc_opt}
h_{\SC}^{\text{opt}}(x) := \left(\frac{1}{d}, \dots, \frac{1}{d}\right)^{\text{T}}. 
\end{align}

We now provide a proof for  \eqref{eq:stochastic_finalexp}. Note that $f(x)$ is a $d$-dimensional probability vector. Since the diagonal part of a pure quantum state forms a probability vector, a random $f(x)$ can be generated by the diagonal part of a pure state in a $d$-dimensional Hilbert space, sampled with respect to the Haar measure. Then, the averaging over all such maps $f$  can be performed by integrating over the Haar measure on the state space. Let $\ket{\psi}$ denote a pure state in a $d$-dimensional Hilbert space $\HC_d$, and let us consider the following expansion 
$
    \ket{\psi} = \sum_{l=1}^d \psi_l \ket{l},
$
where $\psi_l :=  \langle l \vert \psi\rangle$, and where $\ket{l}$ are computational basis states in $\HC_d$. Let $\vert \psi_l \vert^2 = r_l$. Then, the average in \eqref{eq:average_expression} can be computed as 
\begin{align}
\mathbb{E}_f\Vert f(x) - h^{\text{opt}}_{\SC}(x) \Vert^2_1 
&= \int d(\psi) \left( \sum_{l=1}^d \left| \vert \psi_l\vert^2 - \frac{1}{d}   \right| \right)^2\\
& =  \Gamma(d) \int \left( \sum_{l=1}^d \left| r_l- \frac{1}{d}   \right| \right)^2\delta\left(1 - \sum_{j=1}^d r_j\right) \prod_{j=1}^d dr_j \label{eq:sm-stoch-eq2}  \\
& = \Gamma(d) \left[ \int \left( \sum_{l=1}^d \left| r_l- \frac{1}{d}   \right|^2 + \sum_{k\neq l} \left| r_k- \frac{1}{d}   \right|\left| r_l- \frac{1}{d}   \right|  \right)\delta\left(1 - \sum_{j=1}^d r_j\right) \prod_{j=1}^d dr_j \right] \\
    & = \Gamma(d+1) \int  \vert r_1 - 1/d \vert^2 \delta\left(1 - \sum_{j=1}^d r_j\right) \prod_{j=1}^d dr_j \nonumber \\
    & \qquad + (d-1) \Gamma(d+1) \int_0^1 dr_1 \vert r_1 - 1/d \vert \left( \int_0^{\infty} \vert r_2 - 1/d \vert  \delta\left(1 - r_1 - \sum_{j=2}^d r_j\right) \prod_{j=2}^d dr_j\right) \label{eq:two-integrals}.
\end{align}
The first equality holds from the definition of the $1$-norm distance and from \eqref{eq:hsc_opt}. The second equality follows from the representation of the Haar integral over pure states in terms of real parameters \cite{singh2016average}, where $\Gamma(d) = (d-1)!$. The first integral in the third equality is symmetric in $r_l$, and therefore, there are $d$ terms which have the same value, and which result in the first integral in \eqref{eq:two-integrals}. Similarly, the second integral in the third equality is symmetric with respect to any values of $k$ and $l$, and therefore, there are $d(d-1)$ equal terms which leads to the second integral in \eqref{eq:two-integrals}. 

The two integrals in \eqref{eq:two-integrals} can be derived by using  properties of Dirac-delta and the Heavyside-theta functions. The first integral in \eqref{eq:two-integrals} simplifies as follows:
\begin{align}
    \int  \vert r_1 - 1/d \vert^2 \delta\left(1 - \sum_{j=1}^d r_j\right) \prod_{j=1}^d dr_j \nonumber &= \int_0^1 dr_1 \vert r_1 - 1/d \vert^2 \int_0^{\infty}  \delta\left(1 - \sum_{j=1}^d r_j\right) \prod_{j=2}^d dr_j\\
    & = \int_0^1 \vert r_1 - 1/d\vert^2 \frac{1}{(d-2)!} (1-r_1)^{d-2} \Theta(1-r_1) dr_1\\
    & = \frac{1}{(d-2)!d^2(d+1)}, \label{eq:first-integ}
\end{align}
where $\Theta(x)$ is the Heavyside-theta function.

Similarly the second integral in \eqref{eq:two-integrals} can be simplified as follows: 
\begin{align}
    \int_0^1 dr_1 \vert r_1 - 1/d \vert &\left( \int_0^{\infty} \vert r_2 - 1/d \vert  \delta\left(1 - r_1 - \sum_{j=2}^d r_j\right) \prod_{j=2}^d dr_j\right)\\
    & = \frac{1}{(d-3)!}\int_0^1dr_1\vert r_1-1/d \vert  \int_0^{\infty} \vert r_2 -1/d\vert (1-r_1-r_2)^{d-3} \Theta(1-r_1-r_2)dr_2  \\
    & = \frac{8(d-1)^d+4d(d-2)^d -d^d-8(d-2)^d}{(d-3)!(d-2)(d-1)(d+1)d^{d+2}}. \label{eq:second-integ}
\end{align}

Intuitively we expect that it is impossible to learn an infinite dimensional map with a finite training set. Thus we require taht for finite $t$ and in the limit $d \rightarrow \infty$ that the average risk is maximal, $\mathbb{E}_{\SC}[\mathbb{E}_f[R_f(h_{\SC})]] = 1$. This is ensured by renormalizing the above expression for $F(d)$, to obtain the final bound 
\begin{align}
    \mathbb{E}_{\SC}[\mathbb{E}_f[R_f(h_{\SC})]] \geq \left(1-t/d\right) F(d), 
\end{align}
where $F(d)$ is given by \eqref{eq:stochastic_finalexp}.  

We now provide a brief proof for the optimality of the hypothesis in \eqref{eq:hsc_opt}. Since we have no information about $f(x)$ for $x\notin \SC$ in  \eqref{eq:avg-risk-stoch-eq1}, we assume that  $h_{\SC}(x)$ is a fixed hypothesis for each target function $f$ in the average. Let $h_{\SC}(x) = (\beta_1, \dots, \beta_d)^{\text{T}}$, such that $\beta_j \geq 0$ and $\sum_{j=1}^d \beta_j =1$. Then, from arguments similar to \eqref{eq:sm-stoch-eq2}, we get that 
\begin{align}
    &\mathbb{E}_f\Vert f(x) - h_{\SC}(x) \Vert^2_1 \nonumber \\
    & =  \Gamma(d) \int \left(\sum_{l=1}^d \left| r_l- \beta_l   \right| \right)^2\delta\left(1 - \sum_{j=1}^d r_j\right) \prod_{j=1}^d dr_j\\
    & = \Gamma(d) \left[ \int \left( \sum_{l=1}^d \left| r_l- \beta_l   \right|^2 + \sum_{k\neq l} \left| r_k- \beta_k   \right|\left| r_l- \beta_l   \right|  \right)\delta\left(1 - \sum_{j=1}^d r_j\right) \prod_{j=1}^d dr_j \right] \\
    & = \Gamma(d) \sum_{l=1}^d\int_0^1  \vert r_l - \beta_l \vert^2 \frac{1}{(d-2)!} (1-r_l)^{d-2} \Theta(1-r_l) dr_l \nonumber \\
    & \qquad +  \frac{\Gamma(d)}{(d-3)!} \sum_{k\neq l}\int_0^1 dr_k \vert r_k - \beta_k \vert  \int_0^{\infty} \vert r_l - \beta_l \vert  (1-r_k - r_l)^{d-3} \Theta(1-r_k-r_l) dr_l\\
    & = \sum_{l=1}^d \frac{2+\beta_l(d+1)(\beta_l d-2)}{d(d+1)}+\frac{1}{d(d^2-1)(d-2)}\sum_{k\neq l} \bigg(1+ 4(1-\beta_k-\beta_l)^{d+1}+(d+1)(d\beta_k\beta_l - (\beta_k + \beta_l)) \nonumber \\
    & \qquad \qquad +2(1-\beta_k)^d (-1+\beta_k + \beta_l + d \beta_l) +2(1-\beta_l)^d (-1+\beta_l+\beta_k+d\beta_k)\bigg),\label{eq:optimality-proof-eq1}
\end{align}
where we used arguments similar to those used in deriving  \eqref{eq:first-integ} and \eqref{eq:second-integ}. Finally, by setting the derivative of \eqref{eq:optimality-proof-eq1} with respect to $\beta_l$ equal to zero, we find that $\beta_l = 1/d$ for all $l \in \{1, \dots, d\}$. Moreover, the double derivative of \eqref{eq:optimality-proof-eq1}  with respect to $\beta_l$ is positive at $\beta_l = 1/d$, which proves the optimality of the hypothesis in \eqref{eq:hsc_opt}.

\subsubsection{Classical NFL theorem for bistochastic matrices}

In this section, we derive a NFL theorem for learning bistochastic matrices. A bistochastic matrix is a square matrix such that each entry is non-negative, and each row and column sum to one. Similar to Section \ref{sec:stochastic-nfl}, we formulate the task of learning a bistochastic matrix in terms of learning its columns. Let us assume that $M_{\text{BS}}$ is $d\times d$ bistochastic matrix. Let $\XC$ and $\YC$ denote sets of $d$-dimensional bitstrings, and let $|\XC| = |\YC| = d$. Let $f:\XC \to \mathcal{C}(\YC)$ be a map corresponding to the unknown bistochastic matrix $M_{\text{BS}}$, taking bitstrings from $\XC$ to a convex combination of bitstrings in $\YC$. Then \eqref{eq:tset_stochastic}--\eqref{eq:1norm} also hold for the unknown bistochasic map $f$ and the hypothesis map $h_{\SC}$. Similarly, from the arguments used in deriving \eqref{eq:error_train_test_stochastic}--\eqref{eq:minimization}, we find that 
\begin{align}
\mathbb{E}_f[R_f(h_{\SC})]]  
           \geq\left(1-t/d\right) \left[\mathbb{E}_f\Vert f(x) - h^{\text{opt}}_{\SC}(x) \Vert^2_1\right].\label{eq:average_expression_bi}
\end{align}

We now argue that the optimal hypothesis is given by
\begin{align}\label{eq:OptHypBistoch}
    h_{\SC}^{\text{opt}}(x) := \frac{1}{d-t} \left(v- \sum_{i=1}^t f(x_i)\right),
\end{align}
where $v = \left(1, \dots 1\right)^{\text{T}}$. First note that the action of the bistochastic map $f$ is unknown on $x \notin \SC$. However, as each row in a bistochastic matrix sums to one, some partial information about $f(x)$ can be obtained from  each $x_i \in \SC $. Moreover, since each row of a bistochastic matrix sums to one, the hypothesis matrix should be designed such that sum of its each row should also sum to one. Therefore, we get that
\begin{align}
    \sum_{l=1}^d h^{\text{opt}}_{\SC}(x_l) &= v\\
   \to  \sum_{l=t+1}^d h^{\text{opt}}_{\SC}(x_l) &= v - \sum_{l=1}^t f(x_l)\\
 \to   h_{\SC}^{\text{opt}}(x) &= \frac{1}{d-t} \left(v- \sum_{l=1}^t f(x_l)\right) \label{eq:bistochastic-sm-eqb45},
\end{align}
where in the second equation we used \eqref{eq:perfect_train_stochastic}, and in the third equation we used the fact that the hypothesis map should be the same for all $x_l \notin \SC$ as for those cases $f(x_l)$ is equally unknown. Here, we remark that \eqref{eq:bistochastic-sm-eqb45} is valid for any  $x\notin \SC$.   

To perform the averaging over all bistochastic maps, we first assume that a random bistochastic matrix is generated by sampling a $d\times d$ unitary matrix $U$ with respect to the Haar measure, followed by replacing each matrix element $u_{ij}$ of $U$ with $\vert u_{ij}\vert^2$. This construction leads to a bistochastic matrix as for any unitary matrix $U$, the following holds: $\sum_{i}\vert u_{ij}\vert^2 = \sum_{j} \vert u_{ij}\vert^2=1$. Therefore, the average over all bistochastic maps in \eqref{eq:average_expression_bi} can be calculated as follows: 
\begin{align}
 & \mathbb{E}_f\Vert f(x) - h^{\text{opt}}_{\SC}(x) \Vert^2_1   \nonumber \\
 & = \int d(U) \left(\sum_{i=1}^d  \left|u_{ik}u_{ik}^{*} - \frac{1}{d-t} \left[1 - \sum_{j=1}^t u_{ij}u_{ij}^{*} \right]\right| \right)^2 \label{eq:int_bi_general}\\
 & =  \int d(U) \sum_i\left(u_{ik}u_{ik}^{*} - \frac{1}{d-t} \left[1 - \sum_{j=1}^t u_{ij}u_{ij}^{*} \right] \right)^2 \nonumber \\
 & \qquad \qquad + \int d(U) \sum_{i\neq l} \left|u_{ik}u_{ik}^{*} - \frac{1}{d-t} \left[1 - \sum_{j=1}^t u_{ij}u_{ij}^{*} \right]\right|\left|u_{lk}u_{lk}^{*} - \frac{1}{d-t} \left[1 - \sum_{m=1}^t u_{lm}u_{lm}^{*} \right]\right|\label{eq:integral_bistochastic}\\
 & \geq \int dU \left[\sum_i \left( u_{ik}u_{ik}^{*} - \frac{1}{d-t} \left[1 - \sum_{j=1}^t u_{ij}u_{ij}^{*} \right] \right) \left( u_{ik}u_{ik}^{*} - \frac{1}{d-t} \left[1 - \sum_{m=1}^t u_{im}u_{im}^{*} \right] \right)\right] \\
& = \int dU \Bigg[\sum_{i} u_{ik}u_{ik}u_{ik}^{*}u_{ik}^{*}  - \frac{2}{d-t}\sum_{i}u_{ik}u_{ik}^{*} +\frac{1}{d-t}\sum_{i,m}u_{ik}u_{im}u_{ik}^*u_{im}^* +\left(\sum_{i}\frac{1}{(d-t)^2}\right)\nonumber \\
& \qquad \qquad - \frac{2}{(d-t)^2}\sum_{i,m}u_{im}u_{im}^* +\frac{1}{d-t}\sum_{i,j} u_{ij}u_{ik}u_{ij}^*u_{ik}^* + \frac{1}{(d-t)^2} \sum_{i,j,m} u_{ij}u_{im}u_{ij}^*u_{im}^*  \Bigg] \label{eq:expanded_integral_bis}\\
& = \frac{2}{d+1}-\frac{2}{d-t} + \frac{2t}{(d+1)(d-t)} +  \frac{d}{(d-t)^2} -\frac{2t}{(d-t)^2} + \frac{t(t+1)}{(d+1)(d-t)^2}\label{eq:solution_exp_int_bis} \\
& = \frac{1}{1+d}\left(1 - \frac{1}{d-t}\right), \label{eq:final_int_bis}
\end{align}
where the inequality follows because the second integral in \eqref{eq:integral_bistochastic} is non-negative. In \eqref{eq:expanded_integral_bis}, we used \eqref{eq:sym_int1} and \eqref{eq:sym_int2} to obtain \eqref{eq:solution_exp_int_bis}. Here, we denoted the entries of the column vector $f(x)$ as $u_{ik}u_{ik}^*$, such that $k$ is greater than both $j$ and $m$ in \eqref{eq:integral_bistochastic}. 

Finally, by using \eqref{eq:average_expression_bi} and \eqref{eq:final_int_bis}, we get 
\begin{align}
 \mathbb{E}_{\SC}[\mathbb{E}_f[R_f(h_{\SC})]] \geq    \left(\frac{1}{1+d}\right) \left(1- \frac{t+1}{d} \right). 
\end{align}

We note that the aforementiond bound is not tight as we ignore the second integral in \eqref{eq:integral_bistochastic}. 
However, a tighter bound can be obtained by numerically calculating the average in \eqref{eq:int_bi_general}. To do so, we first generated a random bistochastic matrix by taking the square of the absolute value of each element of a unitary matrix sampled with respect to the Haar measure. For a given number of training pairs, the optimal hypothesis $h_{\SC}^{\text{opt}}$ for each remaining unknown column was calculated using \eqref{eq:OptHypBistoch}. 
Finally, we numerically calculated the square of the $1$-norm distance between  $h_{\SC}^{\text{opt}}(x)$ and $f(x)$ for every $x	\notin \mathcal{S}$. Repeating this process for an ensemble of 1000 random bistochastic matrices allowed us to numerically estimate \eqref{eq:int_bi_general}. The final NFL bound for bistochastic matrices in \eqref{eq:average_expression_bi} was obtained by re-normalising the average distances found numerically to agree with the stochastic bound in the limit of no data (i.e. $t=0$). This renormalisation step is analogous to the renormalisation performed in the stochastic case such that the risk tends to 1 in the limit that $d$ tends to infinity for finite $t$.

\section{Resource requirements to violate the Classical NFL theorems.}\label{sec:ressource}

Here we present expressions for the minimal rank $r$ required to violate the permutation, deterministic, stochastic and bistochastic bounds as a function of the number of training pairs $t$ and the dimension $d$ of the unkown matrix. 

The weakest bound, and therefore the hardest bound to violate, is the bound for invertible deterministic functions (i.e. permutation matrices) specified in \eqref{eq:inverse-func}.  It follows from \eqref{eq:Q-NFL} and \eqref{eq:inverse-func} that the risk after learning a $d$ dimensional unitary using $t$ entangled training pairs of rank $r$ is lower than risk for learning a permutation matrix using $t$ training pairs if 
\begin{align}
    \frac{r^2 t^2 + d+1}{d(d+1)} \geq \frac{t+1}{d},
\end{align}
which implies that 
\begin{align}
    r \geq \sqrt{\frac{d+1}{t}} \, .
\end{align}
Similarly, one can show that to violate the classical bound for $d$ dimensional deterministic matrices the $t$ training pairs must be at least of rank 
\begin{align}
    r \geq \sqrt{\frac{d^2-1}{dt}} \, ,
\end{align}
Finally, to  violate the classical bound for $d$ dimensional stochastic or bistochastic matrices we require 
\begin{align}
    r \geq \sqrt{\frac{d(d+1)}{t^2}\left(1 - \left(1-\frac{t}{d}\right) F(d, t) \right) - \frac{1+d}{t^2}} \, 
\end{align}
where $F(d,t) =F(d)$ is defined in \eqref{eq:stochastic_finalexp} for stochastic matrices and $F(d,t)$ for bistochastic matrices is determined numerically. 

In Fig.~\ref{fig:ClassicalBounds}(b), we plot these bounds for the case of learning an $8\times 8$ dimensional permutation, deterministic, bistochastic and stochastic matrices respectively. For any number of training pairs the rank (i.e. amount of entanglement) required to violate the permutation bound is greater than the rank required to violate the stochastic bound, with the rank required to violate the deterministic and bistochastic bounds sitting between these two extremes. This makes sense from an information theoretic perspective. As remarked at the start of this section, permutation matrices are more constrained than bistochastic and deterministic matrices, which in turn are more constrained than stochastic matrices. As such, permutation matrices are easier to learn classically than stochastic matrices making the classical bound more resource intensive to violate.


\end{document}